\documentclass[12pt,preprint]{aastex}

\slugcomment{submitted to ApJ.}

\shorttitle{}
\shortauthors{Sylvie Th\'eado and Sylvie Vauclair}

\begin{document}

\title{On the coupling between helium settling and rotation-induced
mixing in stellar radiative zones : III- Applications to light elements in Pop I main-sequence stars}

\author{Sylvie Th\'eado and Sylvie Vauclair}
\affil{Laboratoire d'Astrophysique, Observatoire Midi-Pyr\'en\'ees, 14
avenue Edouard Belin, 31400 Toulouse, France}

\begin{abstract}

In the two previous papers of this series, we have discussed the importance of the $\mu$-gradients due to helium settling on rotation-induced mixing, first in an approximate analytical way, second in a 2D numerical simulation. We have found that, for slowly rotating low mass stars, a process of ``creeping paralysis" in which the circulation and the diffusion are nearly frozen may take place below the convective zone. Here we apply this theory to the case of lithium and beryllium in galactic clusters and specially the Hyades. We take into account the rotational braking with rotation velocities adjusted to the present observations. We find that two different cells of meridional circulation appear on the hot side of the ``lithium dip" and that the ``creeping paralysis" process occurs, not directly below the convective zone, but deeper inside the radiative zone, at the top of the second cell. As a consequence, the two cells are disconnected, which may be the basic reason for the lithium increase with effective temperature on this side of the dip. On the cool side, there is just one cell of circulation and the paralysis has not yet set down at the age of the Hyades; the same modelisation accounts nicely for the beryllium observations as well as for the lithium ones.

\end{abstract}


\section{Introduction}
This paper is the third of a series on the coupling between helium settling and meridional circulation in stellar radiative zones. The process of meridional circulation and the way it can slow down diffusion have been studied extensively in the past by many authors (see Vauclair \& Th\'eado (2002), hereafter referred as paper I, for references). From the beginning of the computations of meridional circulation and rotation-induced mixing, the importance of the feed-back effect due to nuclearly-induced $\mu$-gradients was well recognised (see Mestel \& Moss 1986), however the feed-back effect due to the diffusion-induced $\mu$-gradients on the meridional circulation was not included until recently. Vauclair (1999) showed how, in slowly rotating low mass stars, the resulting terms in the computations of the meridional velocity could become of the same order of magnitude as the other terms. Th\'eado \& Vauclair (2001) computed these terms for the case of Pop II stars with the assumption of negligible differential rotation. They claimed that the induced diffusion-circulation coupling could be the reason for the very small dispersion of the lithium observed in these stars. More recently (paper I) the effect of differential rotation was discussed and we showed how it could be introduced in a general context. Applications to different stellar populations are underway. Here we discuss the case of Pop II main sequence stars.

While the $\mu$-terms have definitely to be taken into account in the computations, the details of the coupling process which occurs when these terms becomes of the same order of magnitude as the classical terms remain difficult to handle. In paper I, we introduced an analytical approach of the processes and discussed an approximate solution in a quasi-stationary case. We showed that the coupling between helium settling and rotation-induced mixing can slow down both the mixing and the settling when the $\mu$-terms are of the same order of magnitude as the other terms.
In paper II (Th\'eado \& Vauclair 2002), we presented the results of a 2D numerical simulation of the considered processes which helps visualizing the situation. In the example which has been chosen (slowly rotating low mass stars), the $\mu$-terms become of the same order of magnitude as the classical terms, in a short time scale compared to the main-sequence lifetime, thereby creating a frozen region where the circulation does not proceed anymore. However when this occurs the equilibrium between the opposite meridional currents is permanently destabilized by the helium settling below the convective zone. As a consequence, a new motion develops, which mixes up the zone polluated by diffusion. Below this region the circulation remains nearly frozen. This process acts as a restoring system for the element abundance variations and increases in a significant way their settling time scales. We then suggested that diffusion and mixing react in such a way as to keep both the horizontal and vertical $\mu$-gradients constant in the frozen region, while they proceed freely below.

We now apply this prescription to light elements in population I stars. In this paper, we treat the case of the main-sequence F and G stars. In forthcoming papers we will extend our study to the case of sub-giant F stars and we will discuss the effect of differential rotation in the case of Pop II stars.
We emphasize that up to now these studies have been applied to low mass stars with deep convective zones, in which the Gratton-$\rm{\ddot{O}}$pik term (see eq. 3) is negligible. This is not the case for F stars with effective temperatures larger than about 6700K (hot side of the lithium dip in galactic clusters). We will show that in this case two different loops of meridional circulation may arise and that the freezing effect on the circulation does not take place in the first loop, below the convective zone. It may however have some influence at the boundary between the two zones, as discussed below.

In section 2, we review the observations of lithium abundances and rotation velocities in the Hyades. 

A theoretical discussion about previous studies is given in section 3.

In section 4 we recall some important expressions of the meridional circulation and the rotation induced-mixing. 

In section 5 we present the results of the evolutionary computations. We show the influence of the feed-back effect due to the diffusion-induced $\mu$-gradients on the meridional circulation. We discuss the important difference in the implied physical processes which occur on the hot and cool sides of the lithium dip. 

In section 6, we compare the results of the computations with the lithium and beryllium observations in galactic clusters. We give our conclusions and discuss future work.

\section{Observational constraints}

\subsection{The lithium dip}
Lithium abundance determinations in population I stars show that lithium is strongly depleted in the main-sequence stars located in a narrow range of 300K in effective temperature around 6700K. This feature called the ``lithium dip'' had been observed for the first time in the Hyades by Wallerstein et al. (1965) and much later by Boesgaard \& Tripicco. (1986). Since these first observations many abundance determinations have confirmed the presence of the dip in all galactic clusters older than 10$^8$ yrs (NGC 6475 (0.2 10$^9$ yrs), Ursa Mayor (0.5 10$^9$ yrs), Praesepe (0.8 10$^9$ yrs), Coma Berenice (0.8 10$^9$ yrs), NGC 752 (2.2 10$^9$ yrs)) as well as in field stars. In younger clusters (Alpha Persei (0.05 10$^9$ yrs), Pleiades (0.08 10$^9$ yrs)), the phenomenon is not observed, although a small decrease in lithium abundances is observed in the Pleiades around 6700K. In clusters older than NGC 752 (M67 (5 10$^9$ yrs), NGC 188 (6.5 10$^9$ yrs)), the stars of the lithium dip already evolved off the main-sequence.  

On the red side of the dip, the lithium abundances increase for smaller effective temperatures until they reach a  sort of plateau, which extends from 6400 to 6000K. In cooler stars (G type stars) the lithium abundances decrease with effective temperature. 

On the blue side of the dip, the lithium abundances increase with the effective temperature. Few abundance determinations are available for main-sequence cluster stars on this side of the dip because most of these stars are fast rotators whereas lithium can be observed in slow rotators only. For T$_{eff}$ greater than 6900 K, the lithium abundances seem to remain constant, close to the galactic value (Log N(Li) $\simeq$ 3.31 with Log N(H)=12); only few Am stars display anomalous abundances.

Lithium observations in the subgiant stars which originate from the hot side of the dip bring some interesting constraints on hydrodynamical processes which occur during the main-sequence phase. The convective envelopes of the subgiants and giants have deepened and dredged-up material from the interior of the stars and thus provide a view of the internal composition. Many observations show that subgiant stars originating from this side of the dip present large range of lithium abundances which cannot be explained by standard dilution alone (Alschuler 1975, Brown et al. 1989, Balachandran 1990, Dias et al. 1999). These observations suggest that a lithium depletion process takes place in the radiative interior during the main-sequence phase even if its effects are not visible at the surface at the age of the Hyades (Vauclair 1991, Charbonnel \& Vauclair 1992).

\subsection{Rotational velocities in the Hyades}
Boesgaard (1987) noticed that the lithium dip coincides in the Hyades with a sharp drop in rotational velocities. The determination of rotational velocities in the Hyades (Gaig\'e 1993), Praesepe and Coma Berenice clusters shows that stars cooler than 6000K are slow rotators (excepted a few M stars). The increase of the rotational velocity coincides with the lithium decrease on the red side of the dip. On the blue side, the influence of the velocity is more difficult to determine because of the observational bias previously mentioned : in hottest stars ($T_{eff}> 7000$ K), the lithium line is very weak and is detectable only in stars whose rotational velocities are small (V$_{rot} < 50$ km.s$^{-1}$). When the effective temperature decreases, lithium can be observed in stars with higher rotational velocities. This observational bias induces a second bias since slow rotator with $T_{eff}>$ 7000K are chemically peculiar stars.

\section{Previous theoretical studies}
Many explanations have already been proposed to account for the observed lithium abundances in the Hyades. Most of them rely on the fact that the lithium dip appears in the special range of effective temperature where there is a transition between stars with deep convective zones (on the cool side) and shallow ones (on the hot side). We will here briefly summarize them. 

The earliest and simplest explanation for the lithium dip was the microscopic diffusion model of Michaud (1986) (see also Proffitt \& Michaud (1991), Richer, Michaud and Proffitt (1992) and Richer \& Michaud (1993) for more sophisticated treatments of the diffusion). In the temperature range of the dip the convective depth rapidly decreases with increasing effective temperature which strongly affects the microscopic diffusion. They predicted that lithium would settle gravitationally out of the convective zone in stars with T$_{eff}<$ 6900K. At hotter temperatures, upward radiative acceleration was predicted to dominate over gravitation, leading to surface lithium enrichment which are not observed. To explain the lack of lithium overabundances, Michaud et al. invoked mass loss. The rise of the lithium abundance on the red side of the dip was explained by the increase of the convective zone depth : in stars cooler than 6400K, it leads to longer diffusion time scales and thus to slower surface lithium depletion. They finally succeeded in reproducing the lithium dip by adjusting two free parameters : the mixing length and the mass loss rate. 

However several observational facts argue against the pure microscopic diffusion hypothesis. First the predicted dip is narrower than observed (Balachandran 1995). Second the carbon, oxygen and boron under-abundances expected in the case of pure diffusion are not observed in the Hyades (Boesgaard 1989, Friel \& Boesgaard 1990, Garc$\rm{\acute{\i}}$a L\'opez et al. 1993). Finally, according to the models, gravitationally settled Li is supported in a region just below the convective zone and is dredged up to the surface as soon as the star evolves off the main-sequence. Observations on M67 (Balachandran 1995) show that this lithium dredge-up is not observed at the turn-off or on the sub-giant branch.

Based on the rough coincidence between the instability strip and the position of the Li dip, Schramm et al. (1990) suggested that pulsation-driven mass loss could account for the lithium dip. This hypothesis is in severe disagreement with several observations published since their study. It would mean that no lithium depletion should be perceived in the mass-losing model until the convective zone reaches the top of the lithium nuclear destruction region, at which time the depletion would be extremely rapid : the observations which indicate that the dip forms gradually contradict this result. Moreover Delyannis \& Pinsonneault (1993) found that beryllium was also depleted in these stars, in contradiction with this theory : according to the mass loss model the lithium should be totally destroyed before Be depletion appears.  

Garc$\rm{\acute{\i}}$a L\'opez et al. (1991) argued that internal waves generated by the convective envelope of a star may be effective in producing a weak mixing in the radiative interior. They succeeded in reproducing the lithium dip but only by increasing the intensity of the waves by a factor 15 above the estimates obtained with straightforward mixing length theory. Moreover their models fail to reproduce the large abundance dispersion on the red side of the dip.

Many theoretical works have also tested the influence of the rotation-induced mixing on the dip stars. Charbonnel et al. (1992) (hereafter CVZ92) studied the lithium depletion by meridional circulation-induced mixing. They used a simplified expression of the meridional velocity which did not take into account the $\mu$-gradients neither the differential rotation. In this case, the meridional velocity vanishes at the depth where $\rho=\Omega^2/ (2 \pi G)$. Below this layer the circulation goes in the classical sense, while it is inverted above : the circulation is divided in two separate cells. CVZ92 pointed out that the separation of the circulation in two cells occurs precisely at the effective temperature of the bottom of the lithium dip: the second cell appears on the hot side while the boundary is inside the convective zone on the cool side. This result helped them give a consistent modelisation of the dip.
Under the effects of the circulation-induced mixing they showed that the theoretical predictions and the observations agreed on the cool side of the dip. For the hot side of the dip they invoked the ``quiet zone'' which separates the two cells of meridional circulation. However it was later pointed out (Michaud, private discussion) that this zone was too thin to prevent microscopic diffusion so that lithium depletion subsisted anyway. 
In the present paper, we will revisit this theory in the new framework of rotation-induced mixing including $\mu$-gradients and differential rotation.

Talon \& Charbonnel (1998) used the lithium observations in the Hyades to assess the role of meridional circulation and shear turbulence in the transport of angular momentum. Rotation-induced mixing was included as described by Zahn (1992). They computed simultaneously the transport of chemicals and the transport of angular momentum due to the wind-driven meridional circulation. 
In this framework, they reproduced the hot side of the lithium dip but not the cool side. As this prescription also fails to reproduce the flat rotation profile in the sun, they deduced that the angular momentum transport is different on both sides of the dip : it would be due to rotation-induced mixing on the blue side, and to another still unknown process on the red side.

When the $\mu$-gradients are not taken into account, the two loops of circulation do not appear on the hot side of the dip. When they are introduced, the two loops reappear (Palacios et al 2002) : this will be discussed again below.

\section{Diffusion and mixing in the presence of $\mu$-gradients}
\subsection{Rotation-induced mixing}
Here we recall some important expressions of the rotation-induced mixing. Following papers I and II, the vertical velocity amplitude of the meridional circulation can be written:

\begin{equation}
U_{r}=\frac{P}{\bar{\rho} \bar{g} \bar{T} C_p (\nabla_{ad}-\nabla+\nabla_{\mu})} \frac{L}{M_*} E_{tot}
\label{Ur}
\end{equation}
with
\begin{equation}
E_{tot}=(E_{\Omega}+E_{\mu}+E_{\zeta}+E_h)
\end{equation}
$\nabla_{ad}$ and $\nabla$ are the usual adiabatic and real ratios  $\displaystyle \left( {d \ln T\over d \ln P }\right)$, $\nabla_{\mu}$ the mean molecular weight contribution $\displaystyle \left( {d \ln \mu  \over d \ln P}
\right)$. The three first terms in $E_{tot}$ represent :
\begin{itemize}
\item the classical Eddington-Sweet term $E_{\Omega}$, 
\end{itemize}
\begin{eqnarray}
\lefteqn{
E_{\Omega} = 2 \left[ 1-\frac{\bar{\Omega}^2}{2 \pi G \bar{\rho}} \right] \frac{\tilde
  g}{\bar g} }\nonumber \\
& &- \frac{\rho_m}{\bar {\rho}} \left\{ \frac{r}{3}\frac{d}{dr} \left[ H_T
\frac{d\zeta}{dr}
-\chi_T\zeta \right]
-\frac{2 H_T}{r} 
  \zeta +\frac{2}{3}\zeta \right\} \nonumber \\
& &- \frac{\bar{\Omega}^2}{2 \pi G \bar{\rho}} \zeta 
\end{eqnarray}
\begin{itemize}
\item the $\mu$-gradient-induced term $E_{\mu}$,
\end{itemize}
\begin{equation}
E_{\mu}=\frac{\rho_m}{\bar{\rho}}\left\{ \frac{r}{3} \frac{d}{dr} \left[ H_T
\frac{d\Lambda}{dr}  - \left(
\chi_{\mu}+\chi_T+1 \right)
\Lambda \right] - \frac{2H_T}{r}\Lambda \right\} 
\label{emu}
\end{equation}
\begin{itemize}
\item the term related to the time variations of the differential rotation $E_{\zeta}$,
\end{itemize}
\begin{equation}
E_{\zeta}=\frac{M_*}{L}\bar{T} C_p \frac{\partial \zeta}{\partial t}
\end{equation}
Here $\zeta$ represents the density fluctuations along a level surface $\displaystyle \frac{\tilde{\rho}}{\bar{\rho}}$ and 
$\displaystyle \Lambda$  refers to the
horizontal $\displaystyle \mu$-fluctuations
$\displaystyle {\tilde{ \mu}\over \overline \mu }$. 

The last term in the expression of $E_{\Omega}$ is derived from the last term of equation (4.20) in Maeder \& Zahn (1998) (hereafter MZ98); in the following of their study, MZ98 chose to neglect it; here this term is kept in the equations because it can become important in stars more massive than 1.3M$_{\odot}$. Note also that in MZ98, $\bar{T}$ which appears here in the expression of $E_{\zeta}$, was forgotten in equation (4.38), as well as in equation (4.36). 

The fourth term in equation (2) has a special behavior. It appears in the case of large horizontal turbulence (see MZ98), due to the fact that it modifies the radiative energy transport. This term can be written :
\begin{equation}
E_h=\frac{6}{r^2} \frac{M_*}{L} C_p \bar{T} D_h \zeta
\label{Eh1}
\end{equation}
or :
\begin{equation}
E_h=\frac{\rho_m}{\bar{\rho}} \frac{2 H_T}{r} \frac{D_h}{K}\zeta
\end{equation}
where K is the thermal diffusivity and D$_h$ the horizontal diffusion coefficient.

In the assumption of a rotation-induced transport of angular momentum, the horizontal diffusion coefficient is computed accordingly, although an unknown coefficient of order one still remains (Talon and Charbonnel 1998). In all cases, $D_h$ depends on $U_r$ so that this term has to be treated in a different way than the three other ones. 

In paper I, we have shown that $\zeta$ can only vary between 0 (no differential rotation) and $\Lambda$ (maximum differential rotation). If $\zeta=0$, the corresponding terms disappear, while for the other extreme, where $\zeta=\Lambda$, we can write : 
\begin{eqnarray}
\lefteqn{
E_{\Omega} = 2 \left[ 1-\frac{\bar{\Omega}^2}{2 \pi G \bar{\rho}} \right] \frac{\tilde
  g}{\bar g}} \nonumber \\
& &- \frac{\rho_m}{\bar {\rho}} \left\{ \frac{r}{3}\frac{d}{dr} \left[ H_T
\frac{d\Lambda}{dr}
-\chi_T\Lambda \right]
-\frac{2 H_T}{r} 
  \Lambda +\frac{2}{3}\Lambda \right\} \nonumber \\
& &- \frac{\bar{\Omega}^2}{2 \pi G \bar{\rho}} \Lambda \\
\lefteqn{
E_{\zeta}=\frac{M_*}{L}\bar{T} C_p \frac{\partial \Lambda}{\partial t}
}\\
\lefteqn{
E_h=\frac{6}{r^2} \frac{M_*}{L} C_p \bar{T} D_h \Lambda}
\end{eqnarray}

In the stationary case the time derivative disappears ($E_{\zeta}$=0) and, according to Chaboyer \& Zahn (1992) :
\begin{equation}
\Lambda=-\frac{r^2}{6 D_h} U_r \frac{\partial{\ln \bar{\mu}}}{\partial r}
\label{lambda}
\end{equation}
In the present paper, the horizontal diffusion coefficient is approximated following MZ98: $D_h=C_h r |U_r|$, where $C_h$ is an unknown coefficient. The absolute value ($|U_r|$) does not appear in MZ98 but it is here necessary since the meridional velocity amplitude can become negative as we will see later.  
Taking into account the parametrisation of $D_h$, we obtain :
\begin{equation}
\Lambda=- \frac{r}{6 C_h} \frac{U_r}{|U_r|}\frac{\partial{\ln \bar{\mu}}}{\partial r}
\label{lambdastat}
\end{equation} 
$U_r$ and $\Lambda$ have logically the same sign. A positive value of $U_r$  (i.e. an upward flow near the rotation axis and a downward flow near the equator), induce molecular weights greater at the rotation axis than at the equator so $\Lambda$ is positive. Inversely a negative value of $U_r$, as it will be the case below the convective zone in the stars more massive than 1.3M$_\odot$, leads to molecular weights smaller at the rotation axis than at the equator and so to negative value of $\Lambda$. 

Using equation \ref{Eh1} (with $\zeta=\Lambda$) and taking into account the expression of $\Lambda$ (eq. \ref{lambda}) leads to :
\begin{equation}
E_h=-\frac{M_*}{L} C_p \bar{T} U_r \frac{\partial \ln \bar{\mu}}{\partial r}
\end{equation} 
$E_h$ is then proportional to $U_r$ while the dependence on the horizontal diffusion coefficient $D_h$ disappears. In other words, as soon as the differential rotation has the maximum possible value ($\zeta = \Lambda$), the results do not depend anymore on the kind of angular momentum transport which occurs in the star.

The $E_h$ term 
must then be introduced in the computations in a different way, on the left side of equation \ref{Ur}. We can write :
\begin{eqnarray}
\frac{\bar{\rho} \bar{g} \bar{T} C_p (\nabla_{ad}-\nabla+\nabla_{\mu})}{P} U_{r}= \frac{L}{M_*} (E_{\Omega}+E_{\mu})
-C_p \bar{T}\frac{\partial \ln \bar{\mu}}{\partial r}  U_{r} 
\end{eqnarray}
and then :
\begin{equation} 
\frac{\bar{\rho} \bar{g} C_p \bar{T}}{P} \biggl(\nabla_{ad}-\nabla+ \nabla_{\mu}+ \frac{P}{\bar{\rho}\bar{g}} \frac{\partial \ln \bar{\mu}}{\partial r} \biggr)U_r
=\frac{L}{M_*} (E_{\Omega}+E_{\mu})
\end{equation} 
Note that $\displaystyle \frac{P}{\bar{\rho}\bar{g}}\frac{\partial \ln \bar{\mu}}{\partial r}=H_p  \frac{\partial \ln \bar{\mu}}{\partial r} =-\nabla_{\mu}$.
The final expression for $U_r$ is :
\begin{equation}
U_{r}=\frac{P}{\bar{\rho} \bar{g} \bar{T} C_p (\nabla_{ad}-\nabla)} \frac{L}{M_*}( E_{\Omega}+E_{\mu})
\label{Ur2}
\end{equation}

The rotational mixing is then introduced in the computations following Zahn (1992) and MZ98 :
\begin{equation}
D_{turb}=D_v+\frac{(r U_r)^2}{30 D_h}
\end{equation}

$D_v$ represents the vertical part of the shear-induced anisotropic turbulence due to the transport of angular momentum. Its efficiency depends on the Richardson criterion (Maeder 1995, Talon and Zahn 1997). Here we chose to parametrize this coefficient in terms of $U_r \cdot r$ as for the other diffusion coefficients, so that we write :
\begin{equation}
D_{turb}=C_v r |U_r|+\frac{(r U_r)^2}{30 D_h}
\end{equation}
or simply :
\begin{equation}
D_{turb}=\alpha_{turb} r |U_r|
\end{equation}
where $\displaystyle \alpha_{turb}=C_v+ \frac{1}{30C_h}$ is an unknown parameter.

\subsection{Microscopic diffusion}
The microscopic diffusion of helium and other elements is computed as described in paper II. At the considered effective temperatures the radiative force on lithium may be important and we have to add it in the lithium diffusion computations. Here we use the approximate formula developped by Michaud et al. (1976) for the radiative accelerations on unsaturated elements : 
\begin{equation}
g_r=1.7 \hspace{0.2cm} 10^8  \frac{T_{e4}^4 R^2}{A T_4 r^2}
\end{equation}
where $T_{e4}=(T_{eff}.10^{-4})$ K, A is the atomic mass of lithium and R the stellar radius.

\section{Computational results}

\subsection{Characteristics of the models}

\begin{table*}[]

\centering

\caption{Main characteristics of the models with $M \leq 1.25 M_{\sun}$.}
\begin{tabular}{cccccc}
\hline\noalign{\smallskip}
 \multicolumn{1}{c}{$M_*/M_{\odot}$}
& \multicolumn{1}{c}{Age}
& \multicolumn{1}{c}{$V_{rot}$}
& \multicolumn{1}{c}{$T_{eff}$}
& \multicolumn{1}{c}{$Li/Li_0$}
& \multicolumn{1}{c}{$Be/Be_0$}
\\
 &(Myrs) &(km.s$^{-1}$) & & & \\
\noalign{\smallskip}
\hline\noalign{\smallskip}
1.0 & 50&14 &5558 &0.570 &0.970 \\
& 100&11 &5563 &0.422 &0.930 \\
& 200&8 &5582 &0.266 &0.868 \\
& 500& 5&5593 &0.120& 0.767 \\
& 800&4 &5651 & 0.107&0.732 \\
\hline\noalign{\smallskip}
1.1 & 50 & 37&5887 &0.987 & 0.999\\
& 100&28 &5937 &0.906 & 0.997\\
& 200&20 &5944 & 0.766& 0.989\\
& 500& 13&5955 & 0.565& 0.959\\
& 800&10 & 5964& 0.470& 0.930\\
\hline\noalign{\smallskip}
1.2 & 50 &62 &6198 &0.980 & 0.998\\
& 100&48 &6225 & 0.843&0.992 \\
& 200&36 &6231 &0.612 & 0.953\\
& 500&24 &6241 & 0.343& 0.825\\
&800&20 & 6249& 0.242& 0.740\\
\hline\noalign{\smallskip}
1.25 &50 &77 &6349 &0.991 &0.997 \\
& 100&64 &6359 &0.812 & 0.985\\
& 200&51 & 6364&0.478 & 0.889\\
& 500& 35& 6371&0.176 & 0.640\\
& 800& 29& 6376& 0.094&0.508 \\
\noalign{\smallskip}
\hline
\end{tabular}
\label{charac}
\end{table*}

\begin{table*}[]
\centering
\caption{Main characteristics of the models with $M \ge 1.3 M_{\odot}$.}
\begin{tabular}{cccccc}
\hline\noalign{\smallskip}
 \multicolumn{1}{c}{$M_*/M_{\odot}$}
& \multicolumn{1}{c}{Age}
& \multicolumn{1}{c}{$V_{rot}$}
& \multicolumn{1}{c}{$T_{eff}$}
& \multicolumn{1}{c}{$Li/Li_0$}
& \multicolumn{1}{c}{$Be/Be_0$}
\\
 &(Myrs) &(km.s$^{-1}$) & & & \\
\noalign{\smallskip}
\hline\noalign{\smallskip}
1.3 &50 &89 &6477 &0.993 &0.996 \\
& 100& 81&6486 & 0.903&0.985 \\
& 200& 70&6492 & 0.571&0.881 \\
& 500& 53&6499 & 0.117&0.480 \\
& 800& 45& 6504&0.034 & 0.291\\
\hline\noalign{\smallskip}
1.35 &50 &94 &6613 &0.978 & 0.996\\
& 100& 89&6620 &0.755 & 0.953\\
& 200& 80&6622 &0.392 & 0.756\\
& 500& 66&6627 &0.050 & 0.296\\
& 800& 58& 6632&0.0086 &0.133 \\
\hline\noalign{\smallskip}
1.4 &50 &97 & 6743&0.993 &0.995 \\
& 100& 94& 6746&0.908 & 0.985\\
& 200& 88& 6748&0.614 & 0.890\\
& 500& 77& 6749&0.144 & 0.450\\
& 800&70 &6750 &0.033 &0.211 \\
\hline\noalign{\smallskip}
1.45 &50 &97 & 6887&0.992 &0.995 \\
& 100&95 &6892 &0.930 & 0.988\\
& 200&91 &6893 &0.710 & 0.917\\
& 500&82 &6888 &0.295 & 0.571\\
& 800& 76&6826 &0.145 &0.349 \\
\hline\noalign{\smallskip}
1.5 &50 &97 &6924 &0.994 & 0.995\\
& 100& 96 &6925 &0.873 &0.983\\
& 200& 93 &6931 & 0.561&0.874\\
& 500& 86 &6914 &0.531 &0.854\\
& 800& 8& 6872&0.531 &0.854 \\
\hline\noalign{\smallskip}
1.55 &50 &100 &7072 &0.993 &0.996 \\
& 100&100 &7077 &0.980 & 0.995\\
& 200&100 &7073 &0.977 & 0.995\\
& 500&100 &7038 &0.924 & 0.979\\
& 800& 100& 6968& 0.917& 0.977\\
\noalign{\smallskip}
\hline
\end{tabular}
\label{charac}
\end{table*}

We have made computations for 10 stellar masses : 1.0 M$_{\odot}$, 1.1 M$_{\odot}$, 1.2 M$_{\odot}$, 1.25 M$_{\odot}$, 1.3 M$_{\odot}$, 1.35 M$_{\odot}$, 1.4 M$_{\odot}$, 1.45 M$_{\odot}$, 1.5 M$_{\odot}$ and 1.55 M$_{\odot}$. Tables 1 and 2 give the main characteristics of these models at different ages, together with the rotation velocities obtained as discussed below.

We use the statistical study of rotation velocities in the Hyades performed by Gaig\'e (1993) in order to estimate the spin down associated to the stars of different masses : we take an initial velocity of 100 km.s$^{-1}$ that corresponds to the mean velocity of hot stars (T$_{eff}>$7000 K). The resulting velocity at the age of the Hyades corresponds to the average values for stars of a given effective temperature and these velocities are obtained by using the Skumanich law to reproduce the spin down of the stars. 

The models are computed using the solar mixing length parameter and the metallicity of the Hyades : [Fe/H]=0.12 (Cayrel de Strobel et al. 1997). In each model, the parameters $C_h$ and $\displaystyle \alpha_{turb}$ (equation 19) are adjusted to obtain the right lithium depletion at the age of the Hyades. It is interesting to note that, while $\displaystyle \alpha_{turb}$ is nearly constant (it slowly increases from 0.15 for 1.5 M$_{\odot}$ stars to 0.3 for 1.1 M$_{\odot}$ stars), the $C_h$ values needed to account for the observations increase from 5 for 1.5 M$_{\odot}$ stars up to 1300 for 1.1 M$_{\odot}$ stars. This is an indication that the horizontal diffusion coefficient must be larger than the values obtained with Zahn 1992 prescription. The case of the 1.0 M$_{\odot}$ star is still different : the values of the coefficients needed to explain the lithium depletion are considerably larger, which means that another process must occur, related to the fact that the bottom of the convective zone is then very close to the lithium destruction layer (e.g. a tachocline mixing). This will be specially discussed in a forthcoming paper (Th\'eado, Richard and Vauclair 2003, hereafter TRV03). 

We will now present separately the results obtained for stars with masses lower or greater than 1.3 M$_{\odot}$. In low mass stars, the $E_{\Omega}$ term keeps the same sign in the whole radiative interior, inducing therefore only one circulation loop. For masses greater than 1.3 M$_{\odot}$, the Gratton-$\rm{\ddot{O}}$pik term included in  $E_{\Omega}$ is negative just below the convective zone and positive in the deep interior. This leads to a change in the sign of $U_r$ and to two different circulation loops. 

\subsection{The case of cool stars (T$_{eff}<$6500K, M$\le$1.25$M_{\odot}$)}
First we present the results obtained for the 4 coolest stars : 1.0 M$_{\odot}$, 1.1 M$_{\odot}$, 1.2 M$_{\odot}$ and 1.25 M$_{\odot}$.
\begin{figure}
\epsscale{0.5}
\plotone{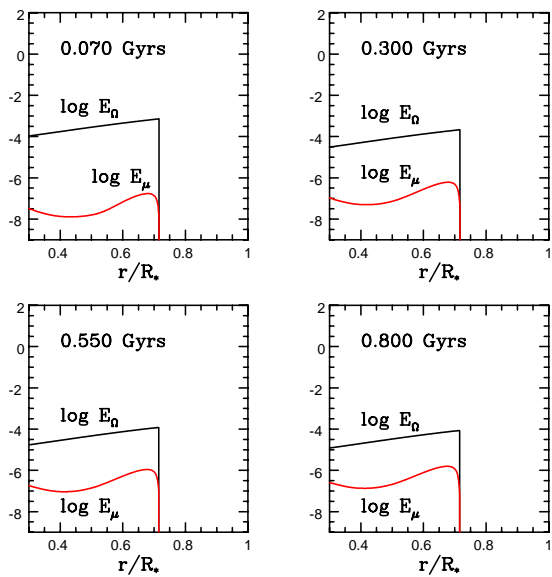}
\caption{Evolution of $\mu$-currents with radius inside a $M=1.0M_{\odot}$ star. The graphs show the variation with depth of both $E_{\Omega}$ and $E_{\mu}$.}
\label{emu1a}
\end{figure}
\begin{figure}
\epsscale{0.5}
\plotone{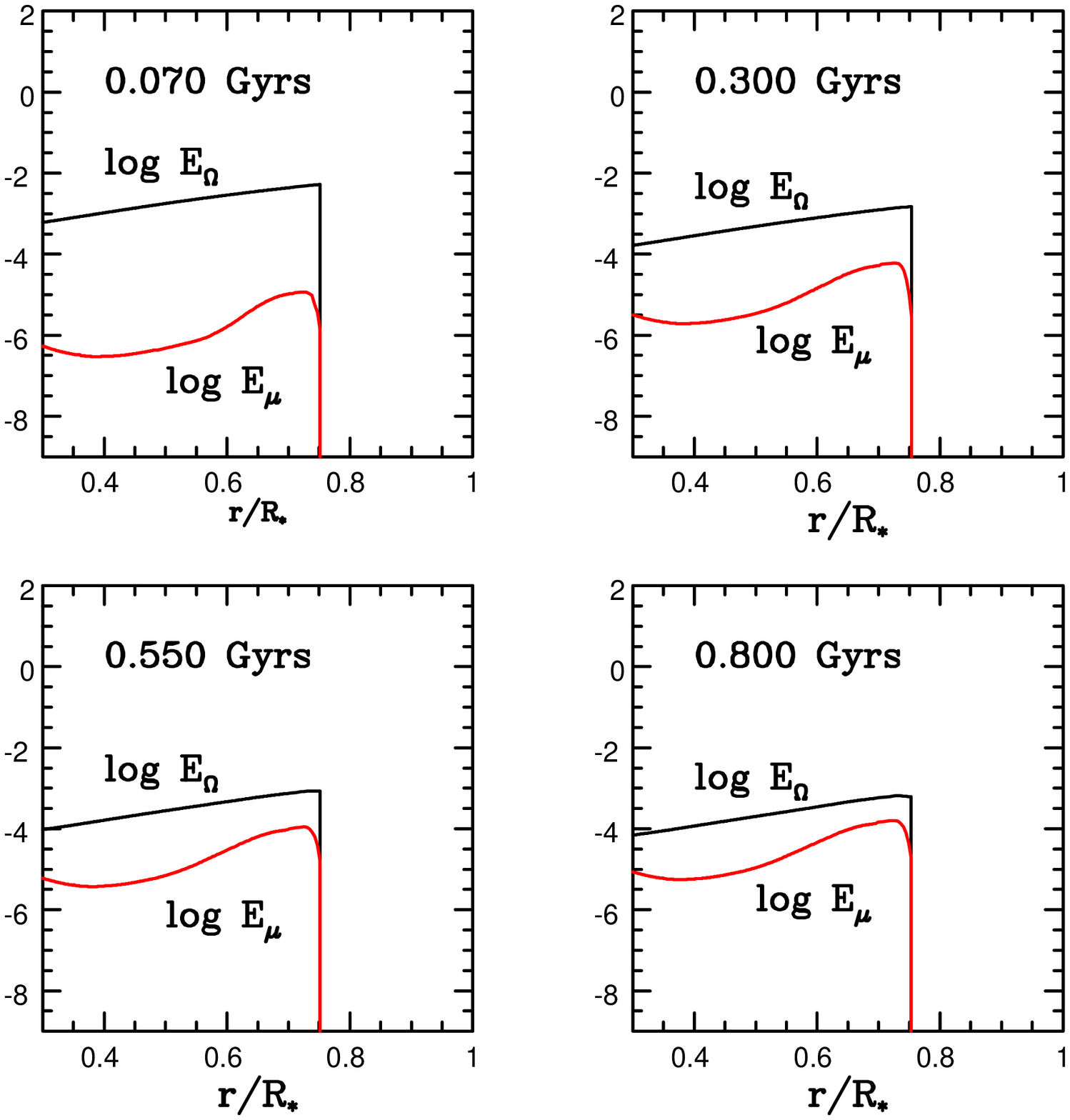}
\caption{Same figures as Fig. \ref{emu1a} for a $M=1.1M_{\odot}$ star.}
\label{emu1b}
\end{figure}
\begin{figure}
\epsscale{0.5}
\plotone{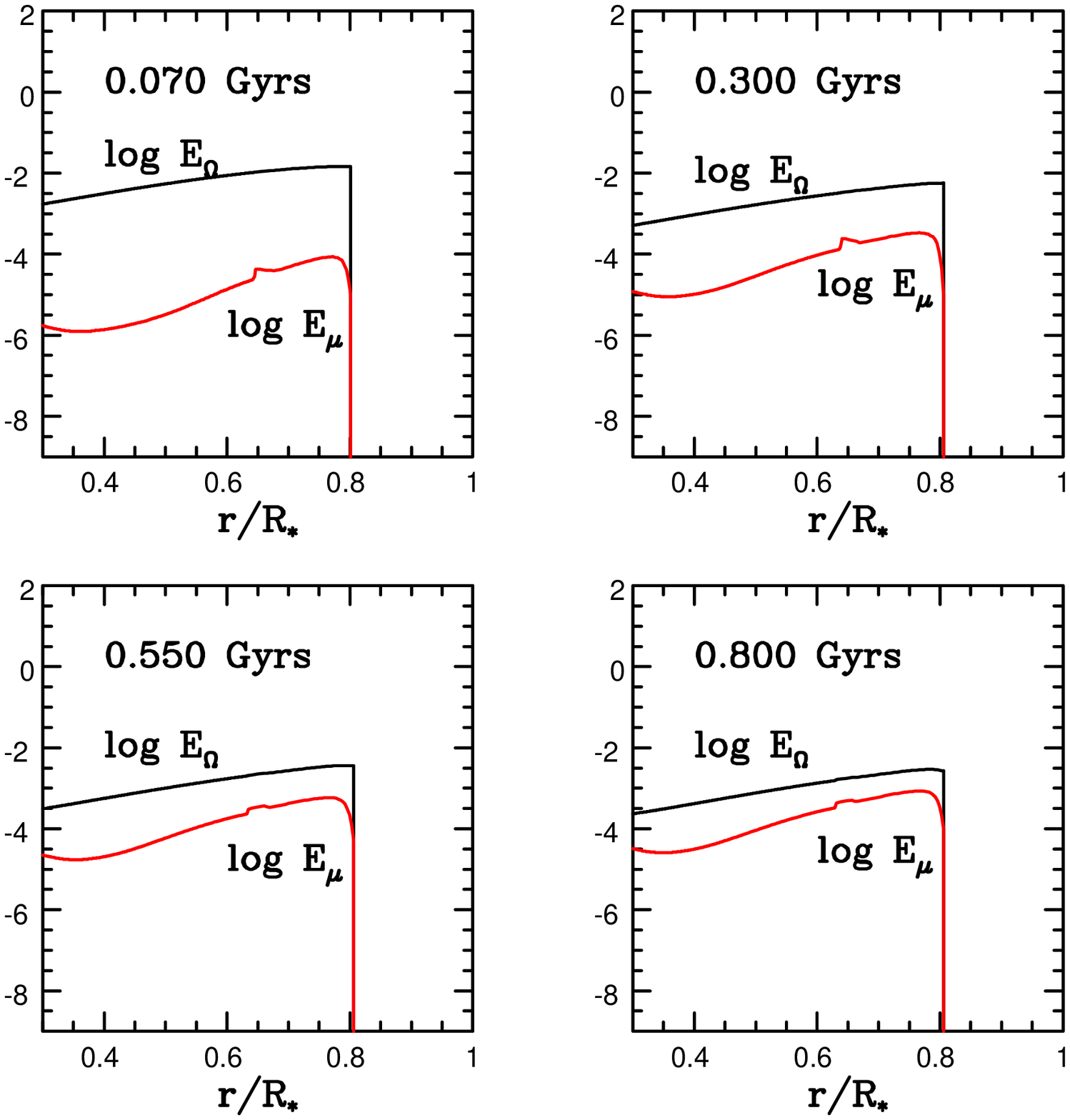}
\caption{Same figures as Fig. \ref{emu1a} for a $M=1.2M_{\odot}$ star.}
\label{emu1c}
\end{figure}
\begin{figure}
\epsscale{0.5}
\plotone{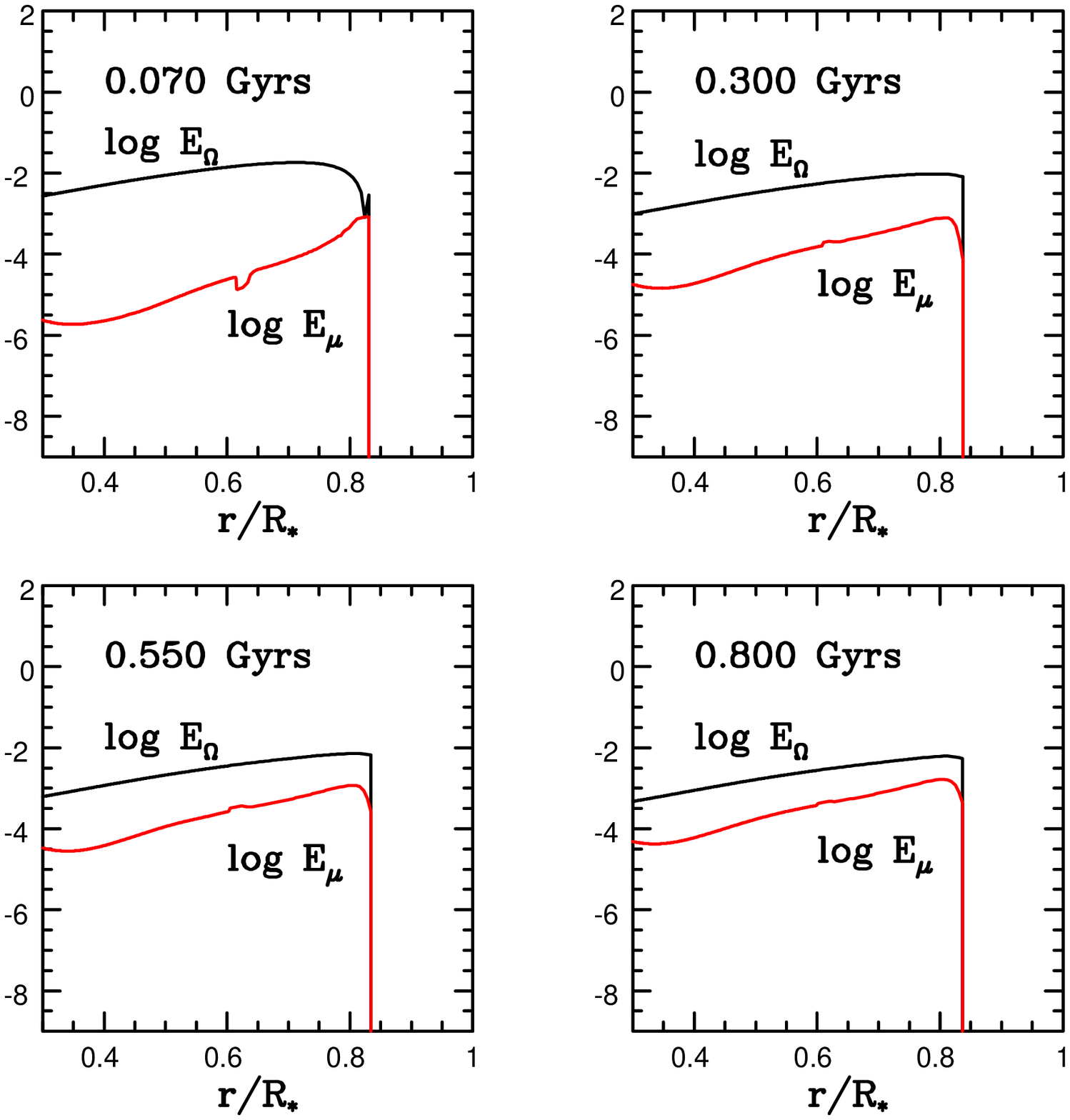}
\caption{Same figures as Fig. \ref{emu1a} for a $M=1.25M_{\odot}$ star.}
\label{emu1d}
\end{figure}

Figure \ref{emu1a} to \ref{emu1d} display the $|E_{\Omega}|$ and $|E_{\mu}|$ profiles below the convective zone at different evolutionary stages. Each star arrives on the main sequence with nearly homogeneous composition. At that time $|E_{\mu}|$ is smaller than $|E_{\Omega}|$ everywhere in the star. Meridional circulation and element diffusion can take place in the stellar radiative region. As a consequence of helium settling the $\mu$-currents increase below the convective zone which reduces the rotational mixing.

The construction of $\mu$-gradients and the increase of $\mu$-currents during the stellar life depends on the competition between several processes. 

i) For increasing effective temperature, the convective depth decreases : the microscopic diffusion time scale consequently decreases which tends to speed up the construction of $\mu$-gradients below the convective zone (since $\Lambda$ is strongly related to the vertical $\mu$-gradients).

ii) For increasing effective temperature, the rotational velocity increases : the $\Omega$-currents are therefore more efficient for increasing masses. This can lead to different effects : a) At the beginning of the stellar life, the advection by circulation which only depends on $\Omega$-currents is more efficient, the $\mu$-gradients are then rapidly built which leads to important $\mu$-currents. These currents reduce the turbulence and the mixing. b) The circulation-induced turbulence is very efficient at the beginning of the stellar life. It tends to homogenize the horizontal layer and to delay the construction of $\mu$-gradients.

In these models, the ``creeping paralysis" which takes place when $|E_{\mu}| = |E_{\Omega}|$ does not occur yet at the age of the Hyades, although it is nearly reached below the convective zone. We will discuss in TRV03 the consequences which appear later in stellar evolution, for the special case of the Sun.

\begin{figure}
\epsscale{1}
\plottwo{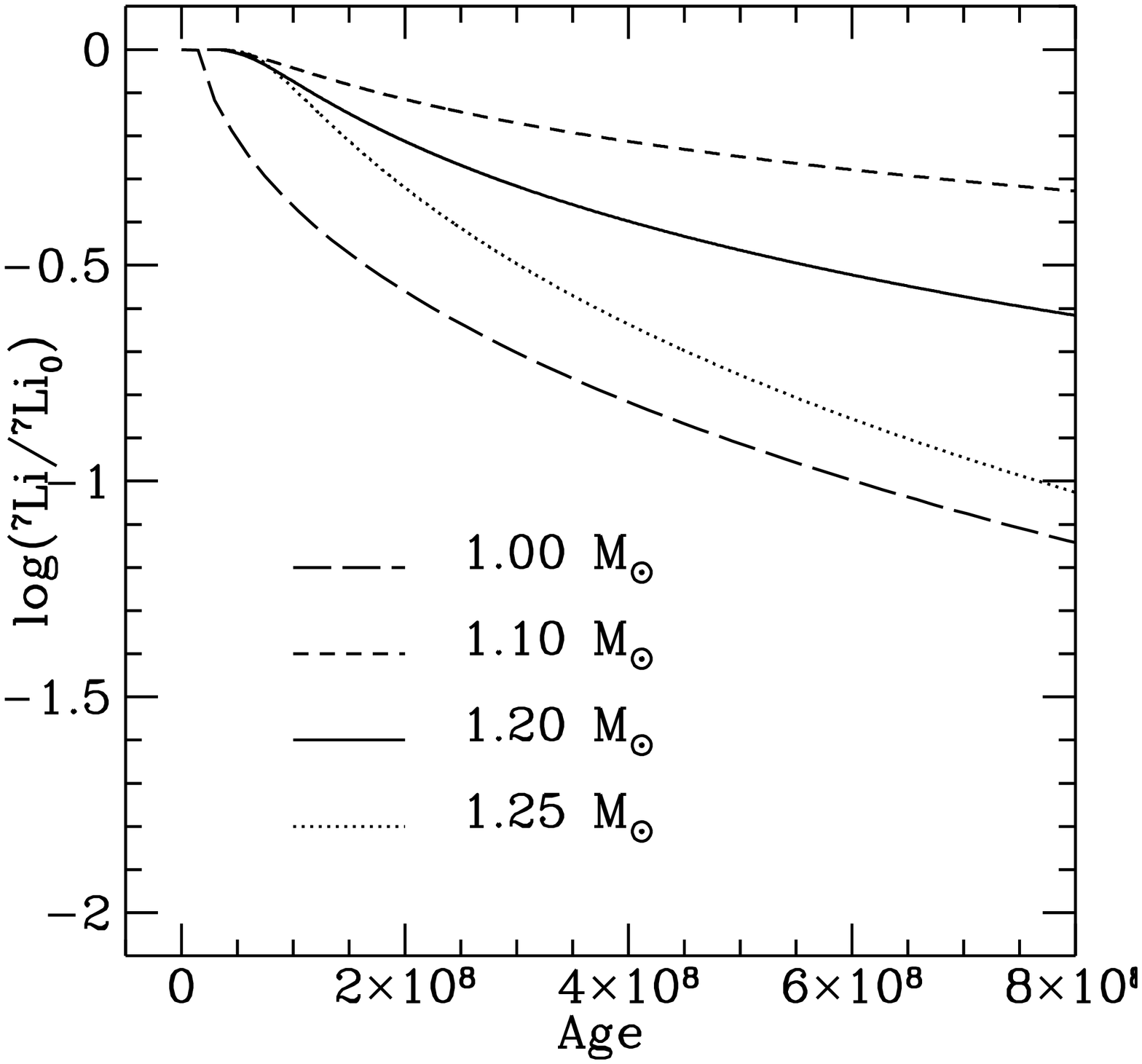}{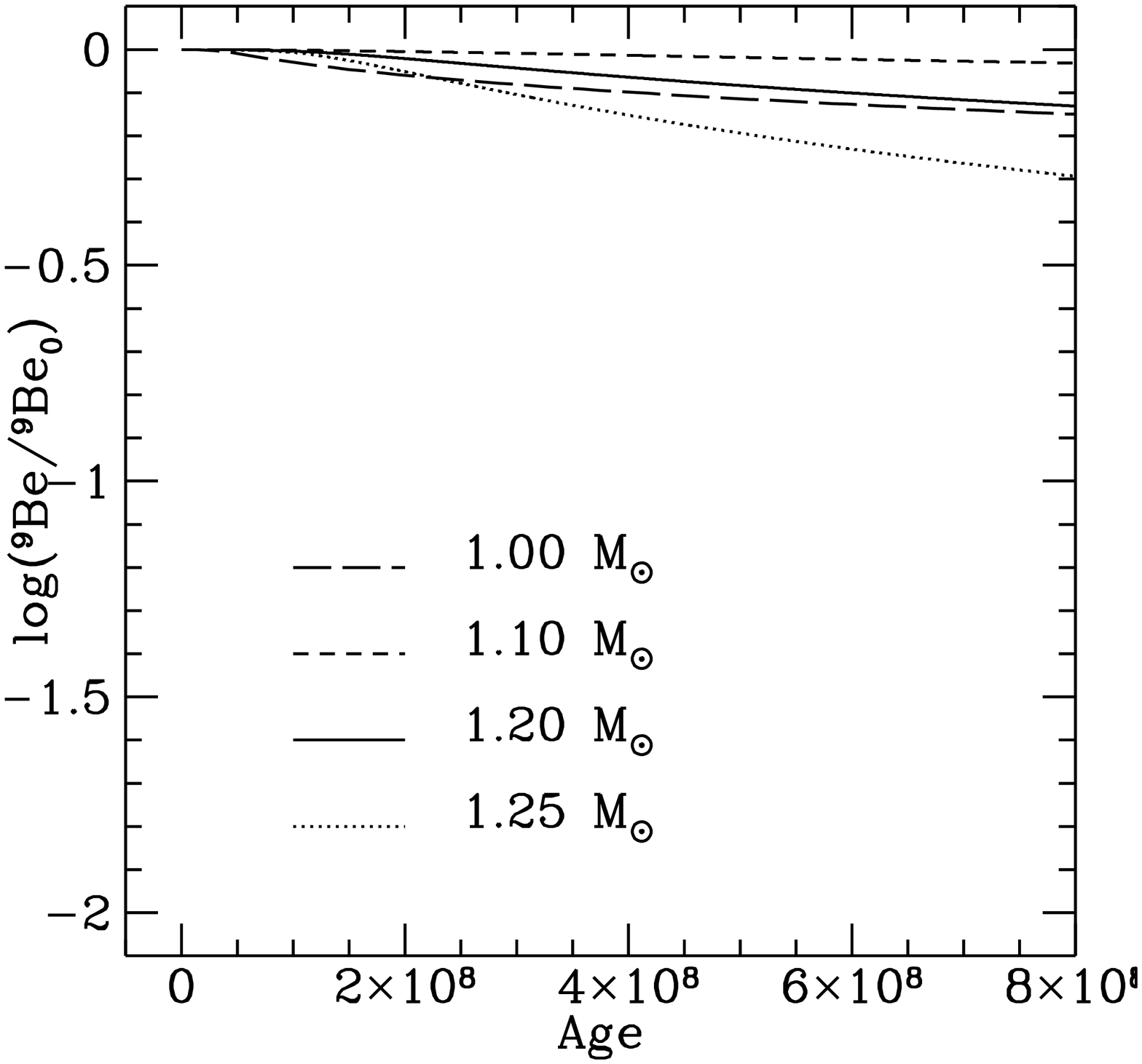}
\caption{Lithium and Beryllium abundance variations at the surface of the stars for masses between 1.0 M$_{\odot}$and 1.25 M$_{\odot}$}
\label{lisurf1}
\end{figure}

Figure \ref{lisurf1} presents the lithium and beryllium abundance variations with time in the outer layers of these four stars. They are the result of the combined effects of diffusion and rotation-induced mixing. We have evoked before the influence of the convective width on microscopic diffusion but the depth of the convective zone plays another important role in the Li and Be depletion rate : the distance between the bottom of the convective zone and the lithium and beryllium nuclear destruction layers decreases with decreasing temperature so these two elements are more easily depleted in the cooler stars than in the hotter.

\begin{figure}
\epsscale{0.6}
\plotone{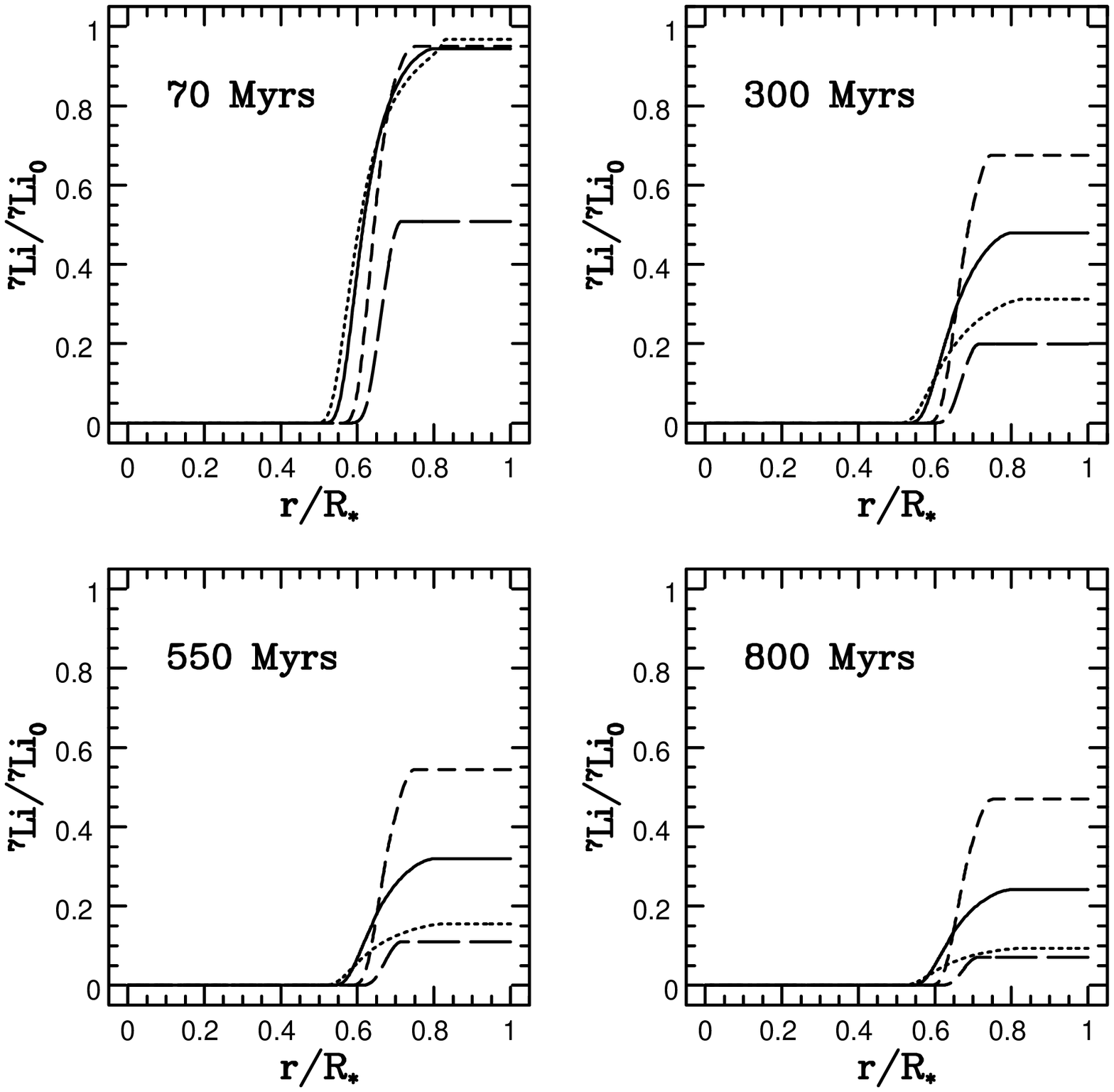}
\caption{Lithium profiles inside the four cooler stars. The long dashed line represents the 1.0M$_{\odot}$ star, the dashed line  : the 1.1M$_{\odot}$ star, the solid line : the 1.2M$_{\odot}$ star and the dotted line : the 1.2M$_{\odot}$ star.}
\label{profilli1}
\end{figure}

\begin{figure}
\epsscale{0.6}
\plotone{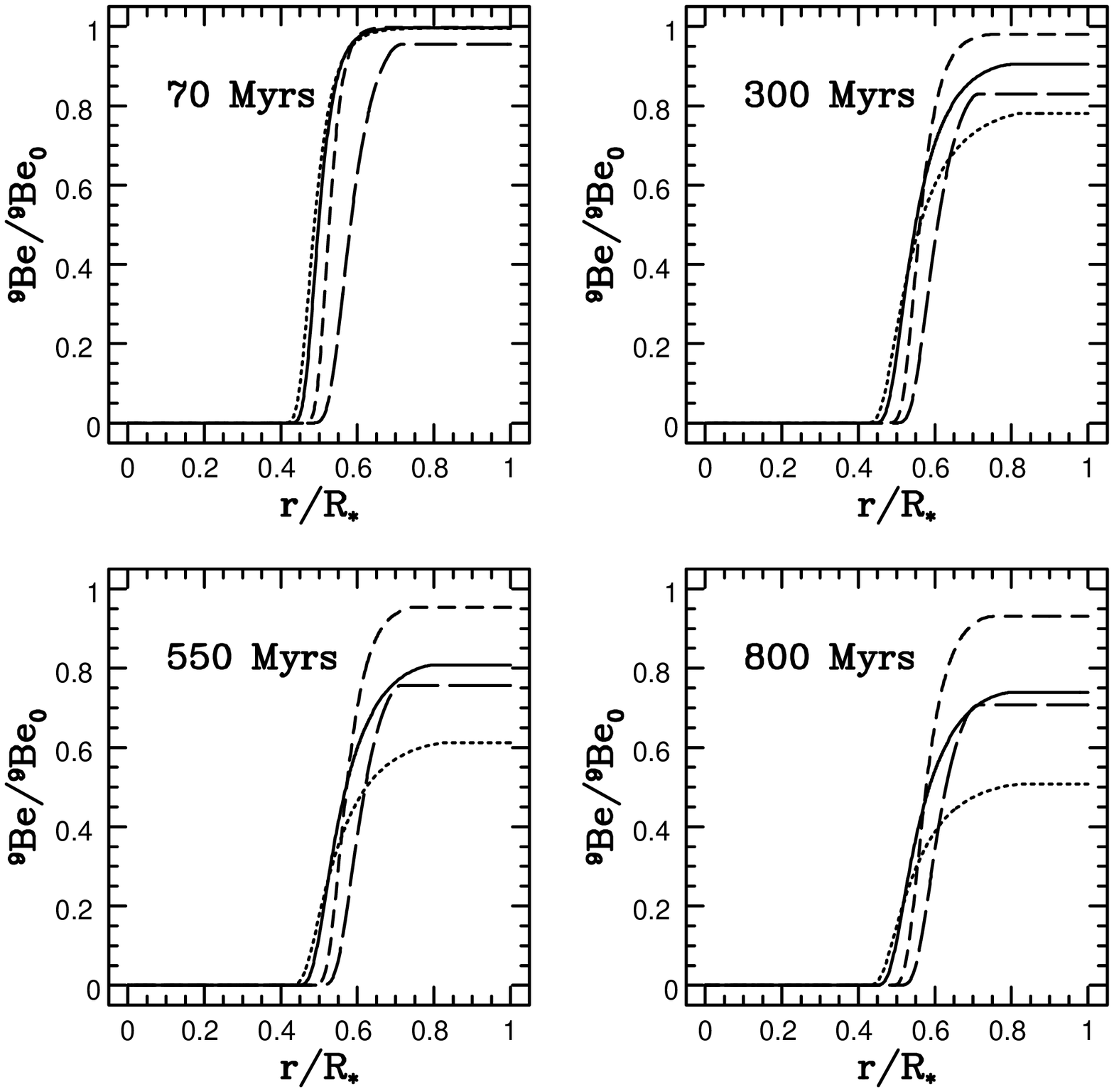}
\caption{Beryllium profiles inside the four cooler stars with the same convention as the lithium profiles figure.}
\label{profilbe1}
\end{figure}

Figure \ref{profilli1} and \ref{profilbe1} displays as an example the lithium profiles inside the 4 stars at different evolutionary stages.

\subsection{The case of hotter stars (T$_{eff}>$6500K, M$\ge$1.3M$_{\odot}$)}
We now present the result obtained for the stars hotter than 6500 K : 1.3 M$_{\odot}$, 1.35 M$_{\odot}$, 1.4 M$_{\odot}$, 1.45 M$_{\odot}$, 1.5 M$_{\odot}$ and 1.5 M$_{\odot}$. In these stars the Gratton-$\rm{\ddot{O}}$pik term included in $E_{\Omega}$ is smaller than one in the deep radiative interior and larger than one below the convective zone. This leads to a change in the sign of $U_r$.
\begin{figure}
\epsscale{0.5}
\plotone{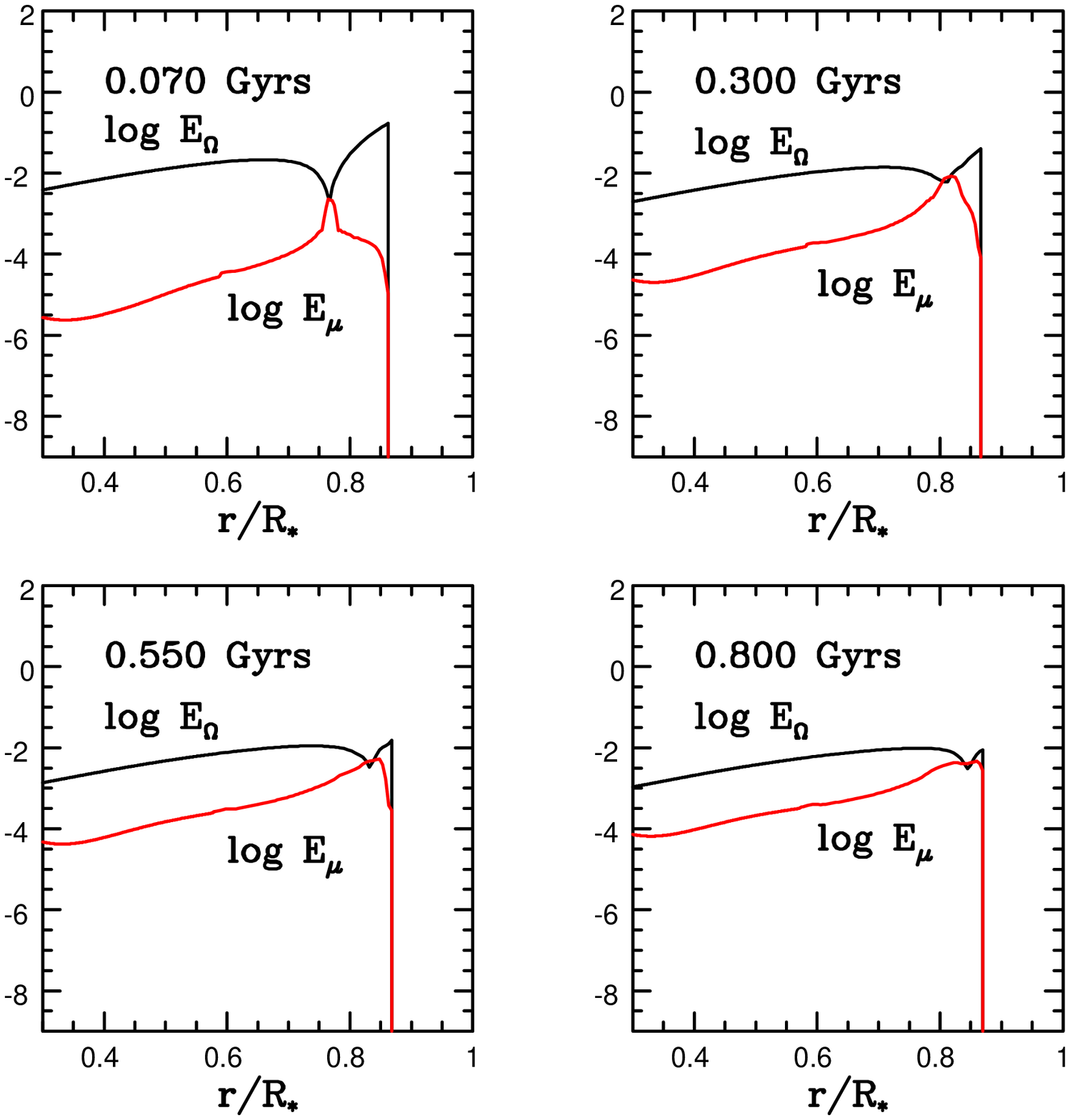}
\caption{Evolution of $\mu$-currents with radius inside a $M=1.3M_{\odot}$ star. The graphs show the variation with depth of both $E_{\Omega}$ and $E_{\mu}$.}
\label{emu2a}
\end{figure}
\begin{figure}
\epsscale{0.5}
\plotone{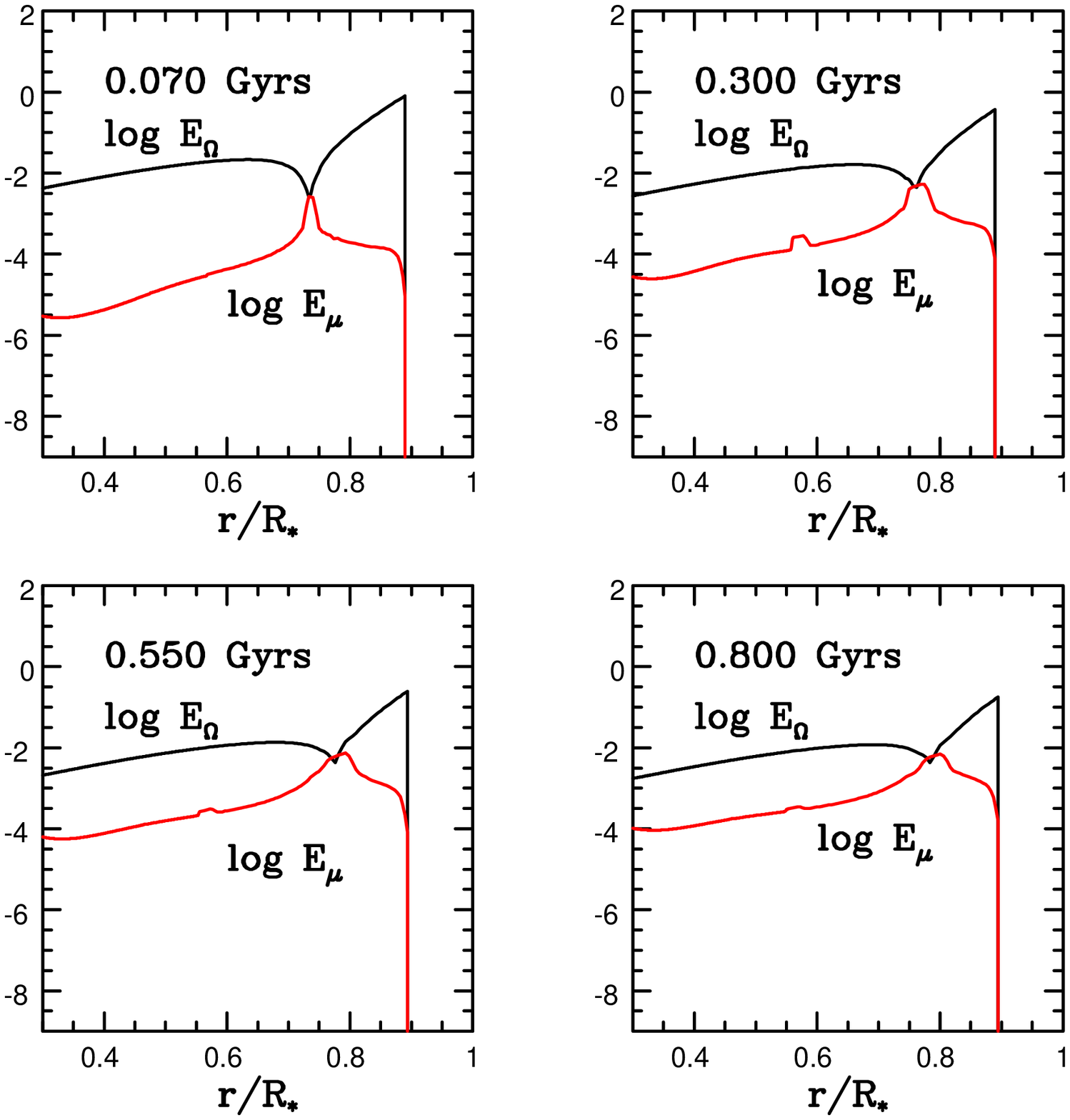}
\caption{Same figures as Fig. \ref{emu2a} for a $M=1.35M_{\odot}$ star.}
\label{emu2b}
\end{figure}
\begin{figure}
\epsscale{0.5}
\plotone{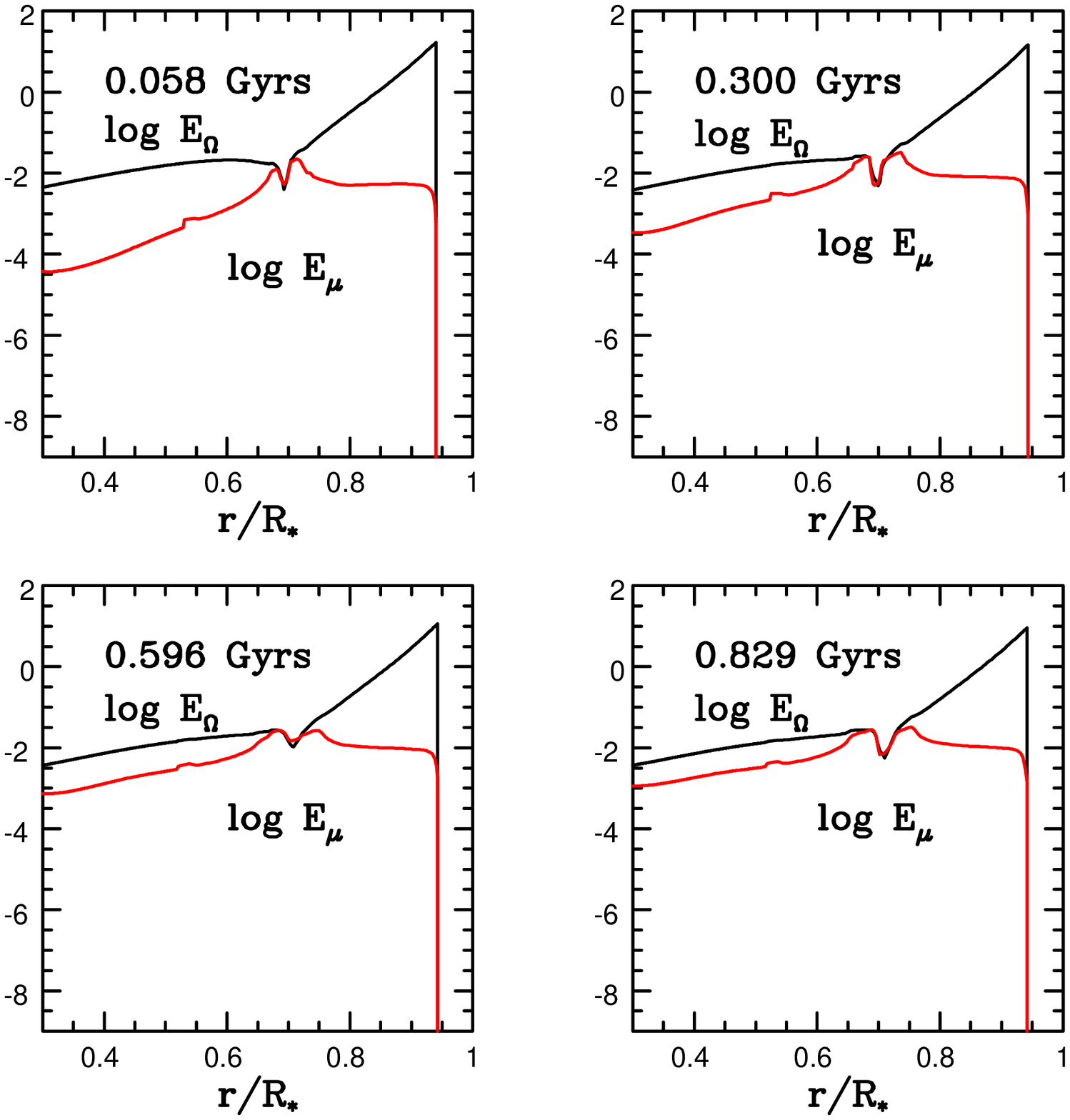}
\caption{Same figures as Fig. \ref{emu2a} for a $M=1.45M_{\odot}$ star.}
\label{emu2c}
\end{figure}
\begin{figure}
\epsscale{0.5}
\plotone{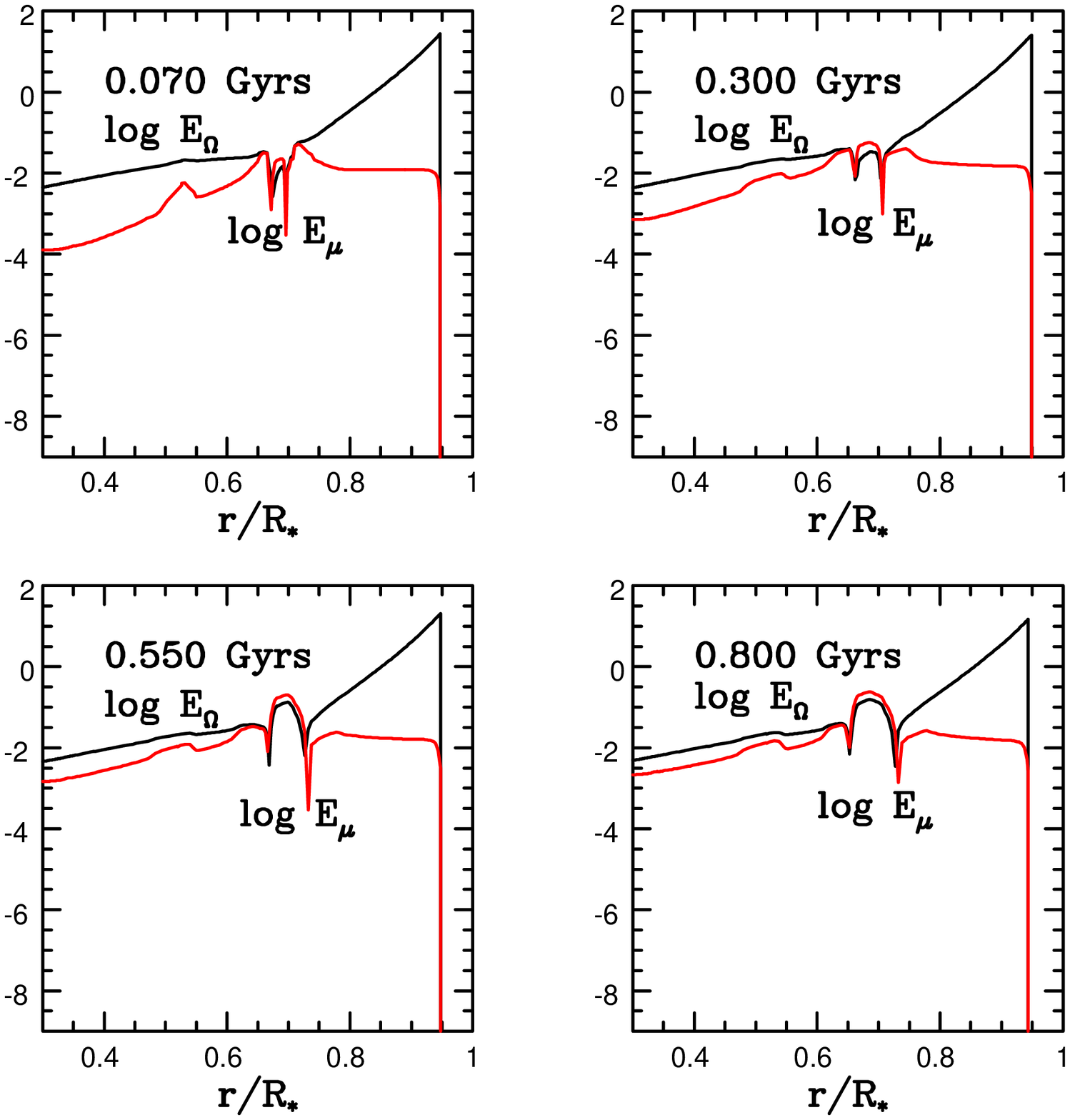}
\caption{Same figures as Fig. \ref{emu2a} for a $M=1.5M_{\odot}$ star.}
\label{emu2d}
\end{figure}

\begin{figure}
\epsscale{0.3}
\plotone{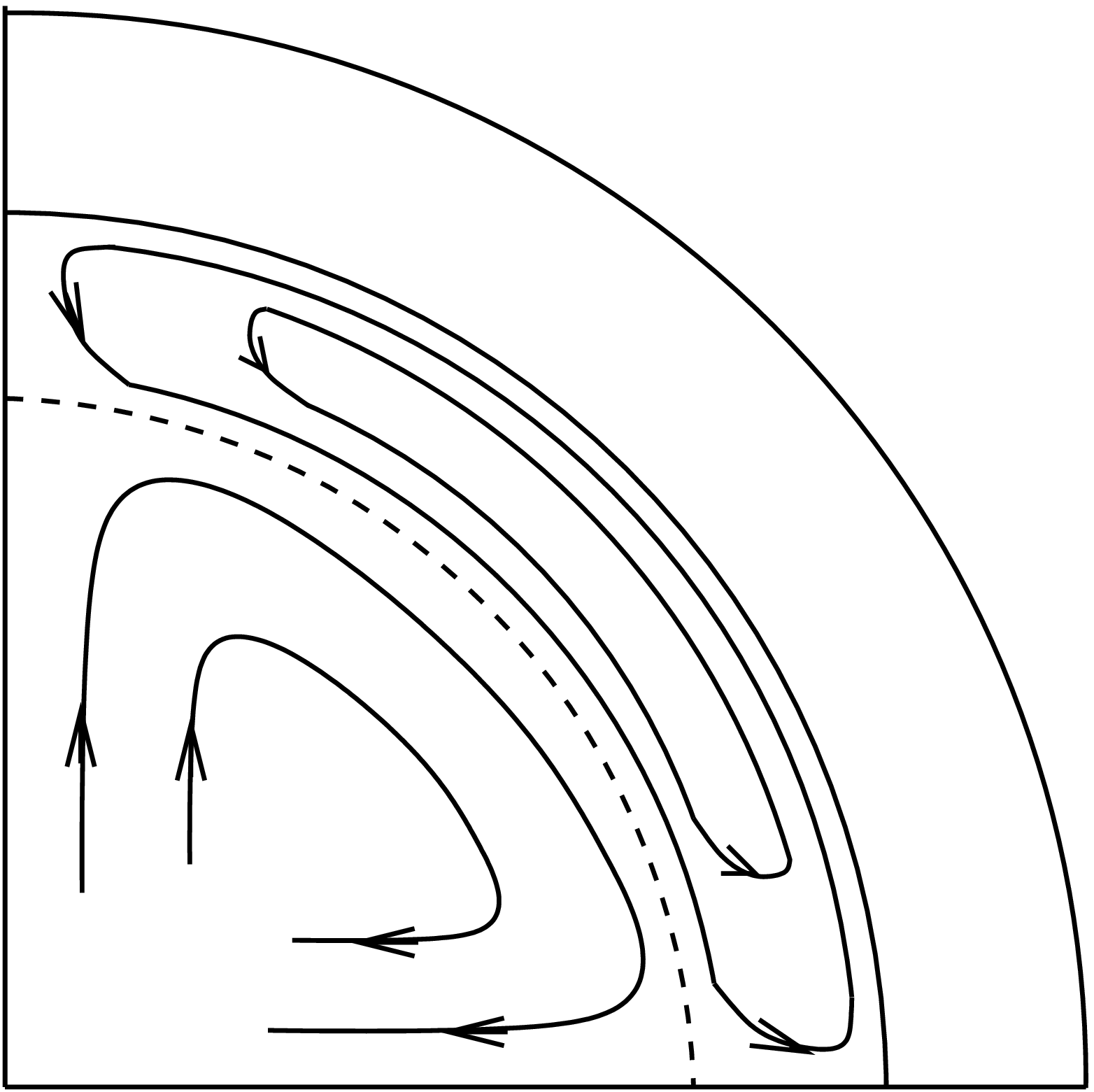}
\caption{Meridional circulation in the stars with M$\geq$1.30M$_{\odot}$. In the less massive stars, the upper circulation loop disappears and the lower loop takes place in the whole radiative interior.}
\label{dboucles}
\end{figure}

Figure \ref{emu2a} to \ref{emu2d} show the $|E_{\Omega}|$ and $|E_{\mu}|$ profiles in four of the considered stars below the convective zone at different evolutionary stages. The drop in the $|E_{\Omega}|$ profiles corresponds to the place where this term changes sign. Two circulation loops then appears : the deeper one goes from the rotation axis to the equator, the upper one goes in the opposite direction from the equator to the rotation axis (see figure \ref{dboucles}). Then the drop in the $|E_{\Omega}|$ profiles also indicates the bottom of the upper circulation loop. 

As said before, in the upper circulation loop, the circulation goes from the equator to the pole. Then it brings $\mu$-enriched matter to the equator and $\mu$-depleted matter to the pole. This leads to negative horizontal $\mu$-gradients ($\Lambda < 0$) and then to $\mu$-currents which go from the pole to the equator (i.e : $E_{\Omega} < 0$ and $E_{\mu} > 0$).

As for cooler stars, each hotter star arrives on the main sequence with nearly homogeneous composition. At that time $|E_{\mu}|$ is smaller than $|E_{\Omega}|$ everywhere in the star. Under the combined effects of meridional circulation and element diffusion the $\mu$-currents increase rapidly. Figure \ref{emu2a} to \ref{emu2d} show that they become of the first order in the transition layer between the two circulation loops. In the 1.3 M$_{\odot}$ and 1.35 M$_{\odot}$ stars, the region where $|E_{\Omega}|\simeq |E_{\mu}|$ is very thin while it becomes significant in the 1.40 M$_{\odot}$ and 1.5 M$_{\odot}$ stars : about 1\% of the radius in the 1.4 M$_{\odot}$ and more than 3\% of the radius in the 1.5 M$_{\odot}$ at the age of the Hyades. 

This leads to the settling in the star of the quasi-equilibrium stage described in paper I and II. The frozen region expands slowly inside the star. Between the base of the convective zone and the top of the frozen region, stellar matter is mixed as well as below the bottom of the frozen region in the radiative interior. The stellar matter in the upper and the lower circulation loops are then disconnected.

The construction of $\mu$-gradients occurs differently here than in cooler stars. A rotational velocity increase leads to more vigourous $\Omega$-currents and then to a more vigourous mixing. In the 1.3 M$_{\odot}$ and 1.35 M$_{\odot}$ stars, the mixing prevents the $\mu$-currents of becoming of the same order of magnitude as $\Omega$-currents in an extended region. In the 1.4 M$_{\odot}$ and 1.5 M$_{\odot}$ stars, the diffusion becomes more efficient than the mixing and forces the $\mu$-currents to become more important. 

The lithium abundance variations with time have been computed using the following scenario : microscopic diffusion damped by rotation-induced mixing occurs normally until the relative difference between $|E_{\mu}|$ and $|E_{\Omega}|$ becomes smaller than 10\%. Then, in the concerned region, mixing is strongly reduced while settling proceeds but with the constraint of keeping constant $\mu$-gradients. Below and above the frozen region, diffusion and mixing are assumed to proceed freely.
\begin{figure}
\epsscale{1}
\plottwo{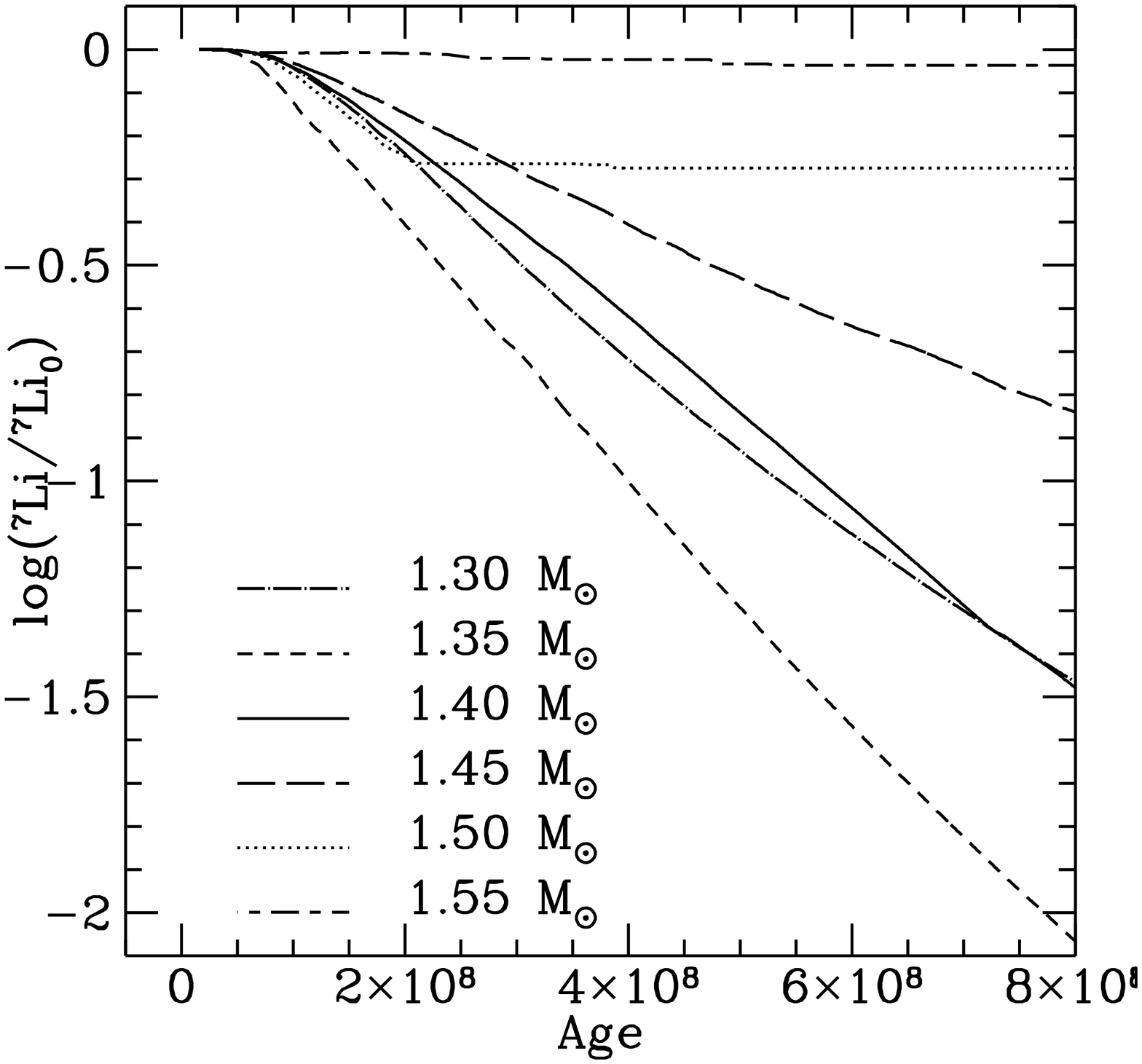}{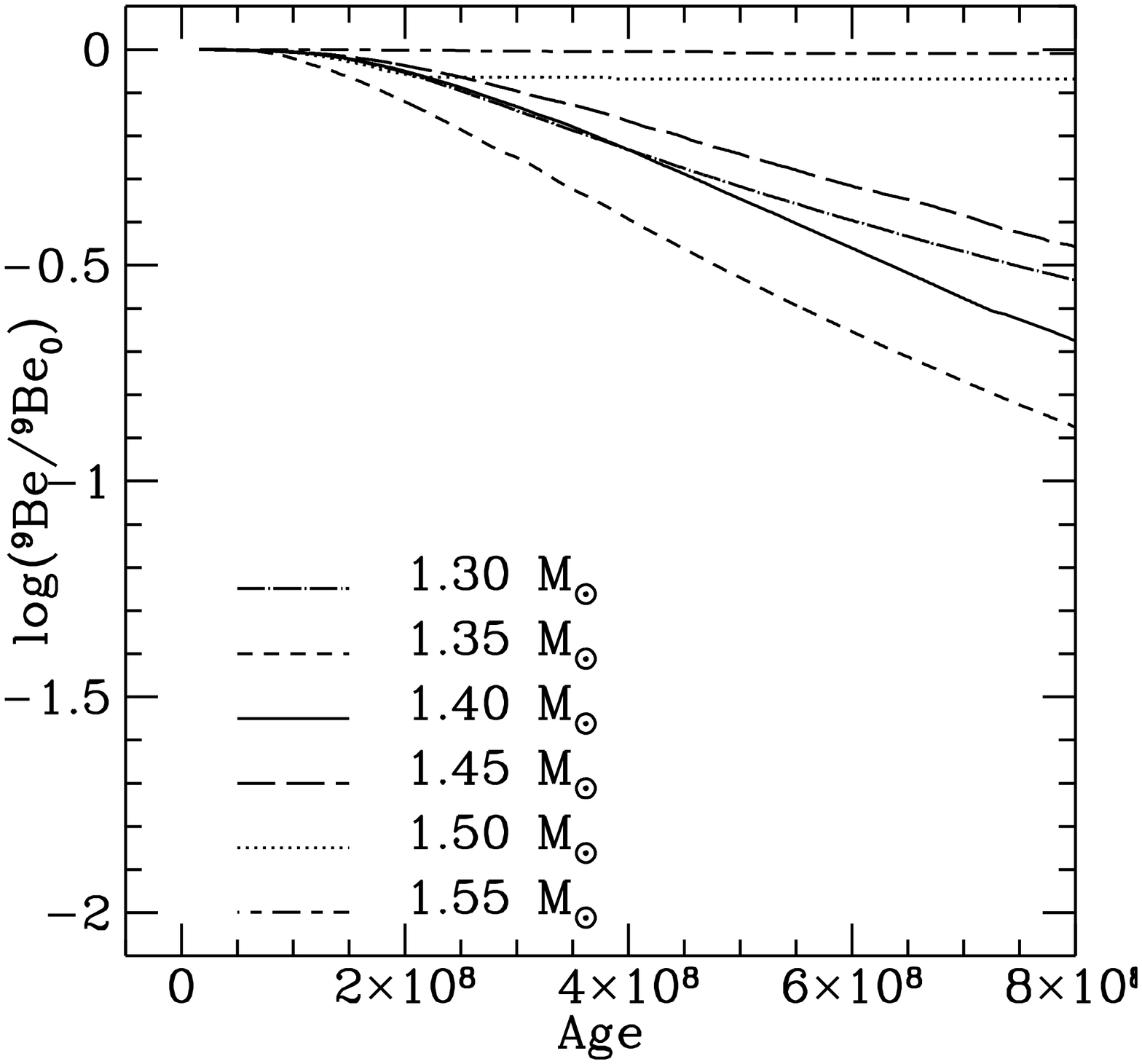}
\caption{Lithium and Beryllium abundance variations at the surface of the stars for masses between 1.3 M$_{\odot}$ and 1.5 M$_{\odot}$}
\label{lisurf2}
\end{figure}

\begin{figure}
\epsscale{0.6}
\plotone{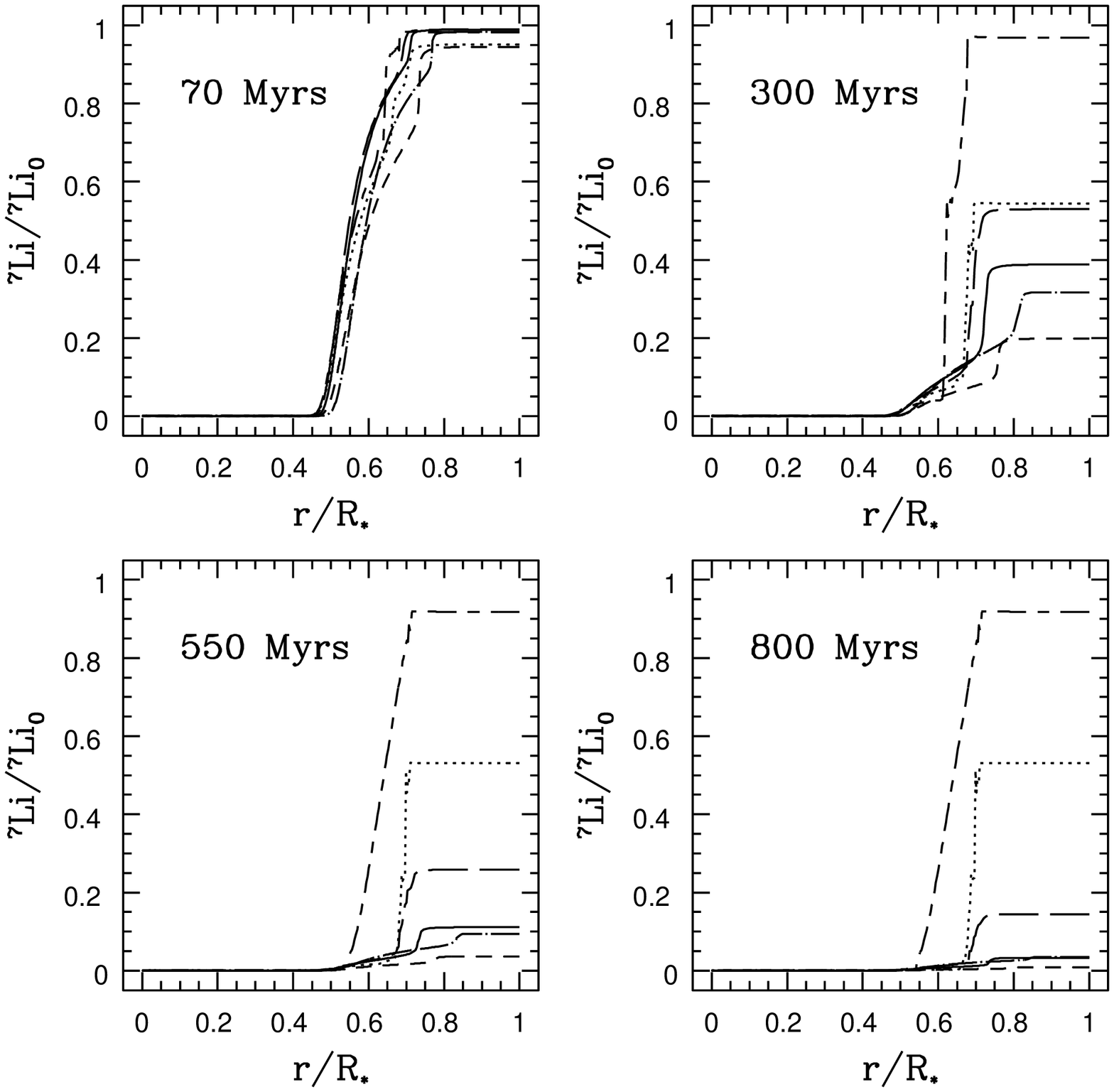}
\caption{Lithium profiles inside the four hotter stars. The long dashed-dotted line represents the 1.3M$_{\odot}$ star, the dashed line  : the 1.35M$_{\odot}$ star, the solid line : the 1.4M$_{\odot}$ star, the long dashed line : the 1.45M$_{\odot}$ star, the dotted line : the 1.5M$_{\odot}$ star and the long-dashed-dashed line : the 1.55M$_{\odot}$ star.}
\label{profilli2}
\end{figure}

\begin{figure}
\epsscale{0.6}
\plotone{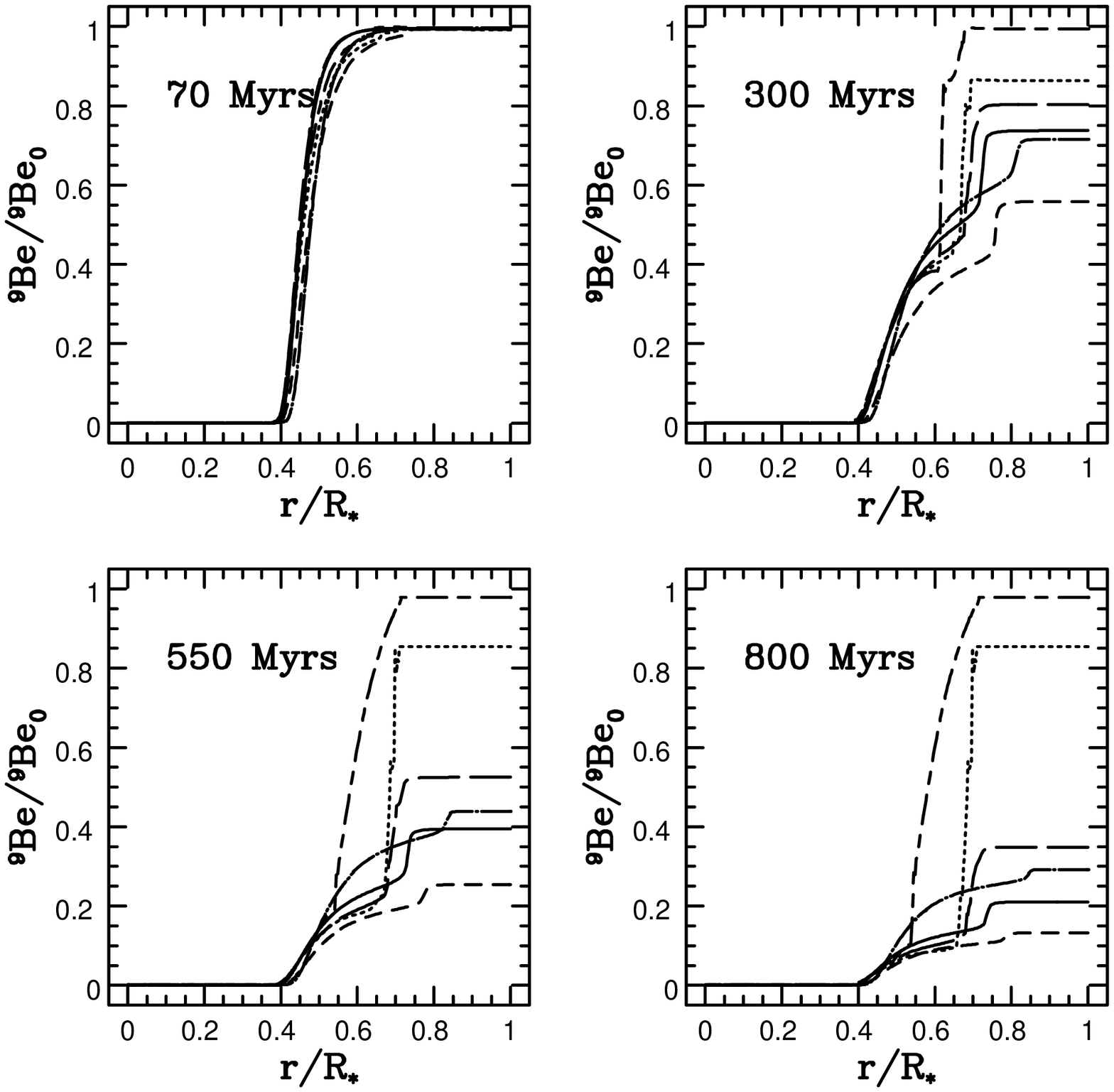}
\caption{Beryllium profiles inside the four hotter stars with the same convention as the lithium profiles figure.}
\label{profilbe2}
\end{figure}

Figure \ref{lisurf2} presents the lithium and beryllium abundance variations with time for the four considered stellar masses. In the 1.3 M$_{\odot}$ and 1.35 M$_{\odot}$ stars, the frozen region is not thick enough to prevent the combined effects of the microscopic diffusion and the mixing : lithium is depleted. In the two other stars, the frozen region is larger and significantly reduces the lithium depletion. The more efficient the diffusion, the more rapid the $\mu$-currents increase. It explains why this process settles down earlier in the 1.5 M$_{\odot}$ star than in the 1.4 M$_{\odot}$ star and why the lithium depletion is less important in the more massive star. Figure \ref{profilli2} and \ref{profilbe2} shows the lithium and beryllium profiles in the star at different evolutionary stages.

\subsection{Discussion} 
\begin{figure}
\epsscale{1.}
\plottwo{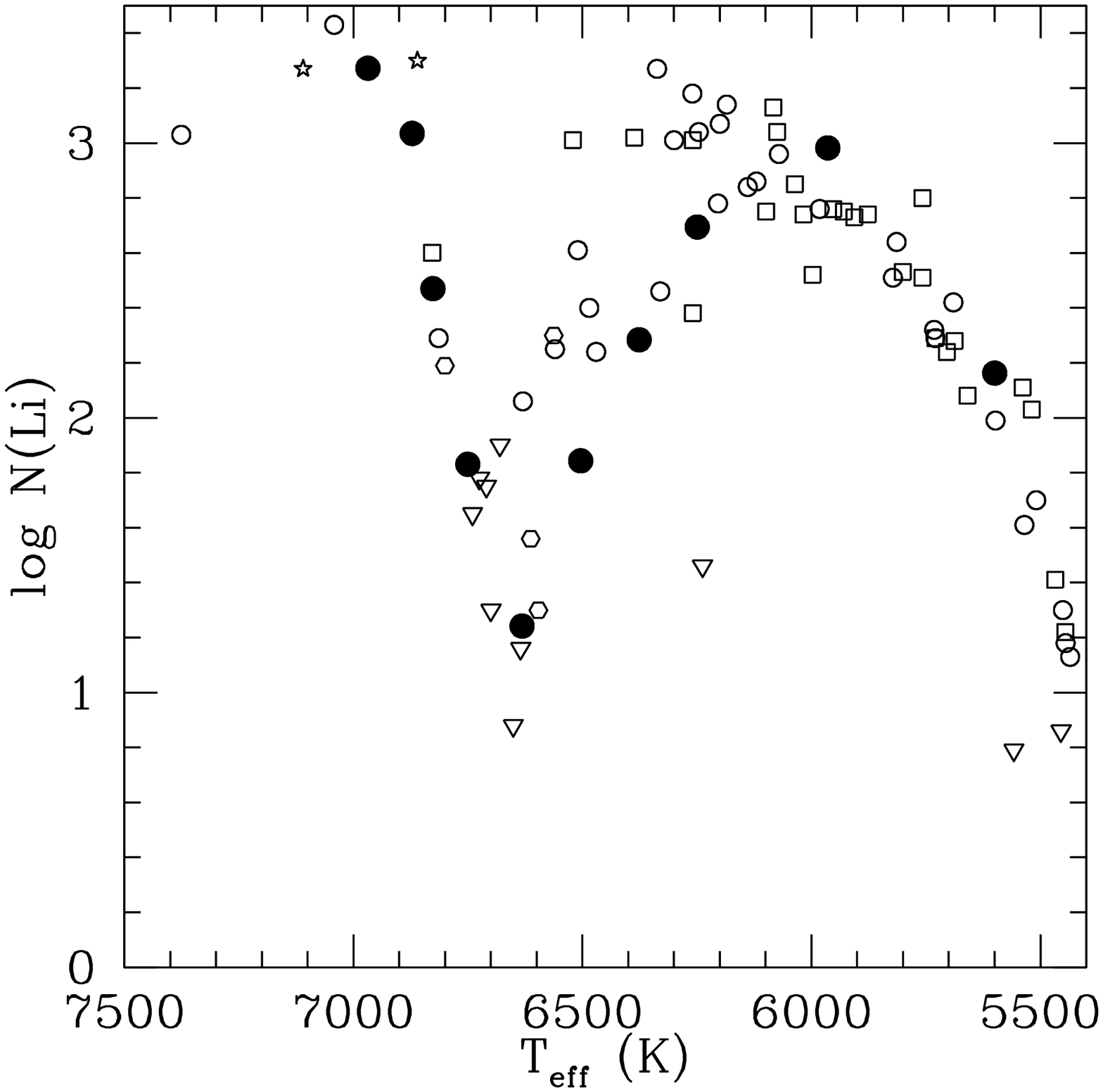}{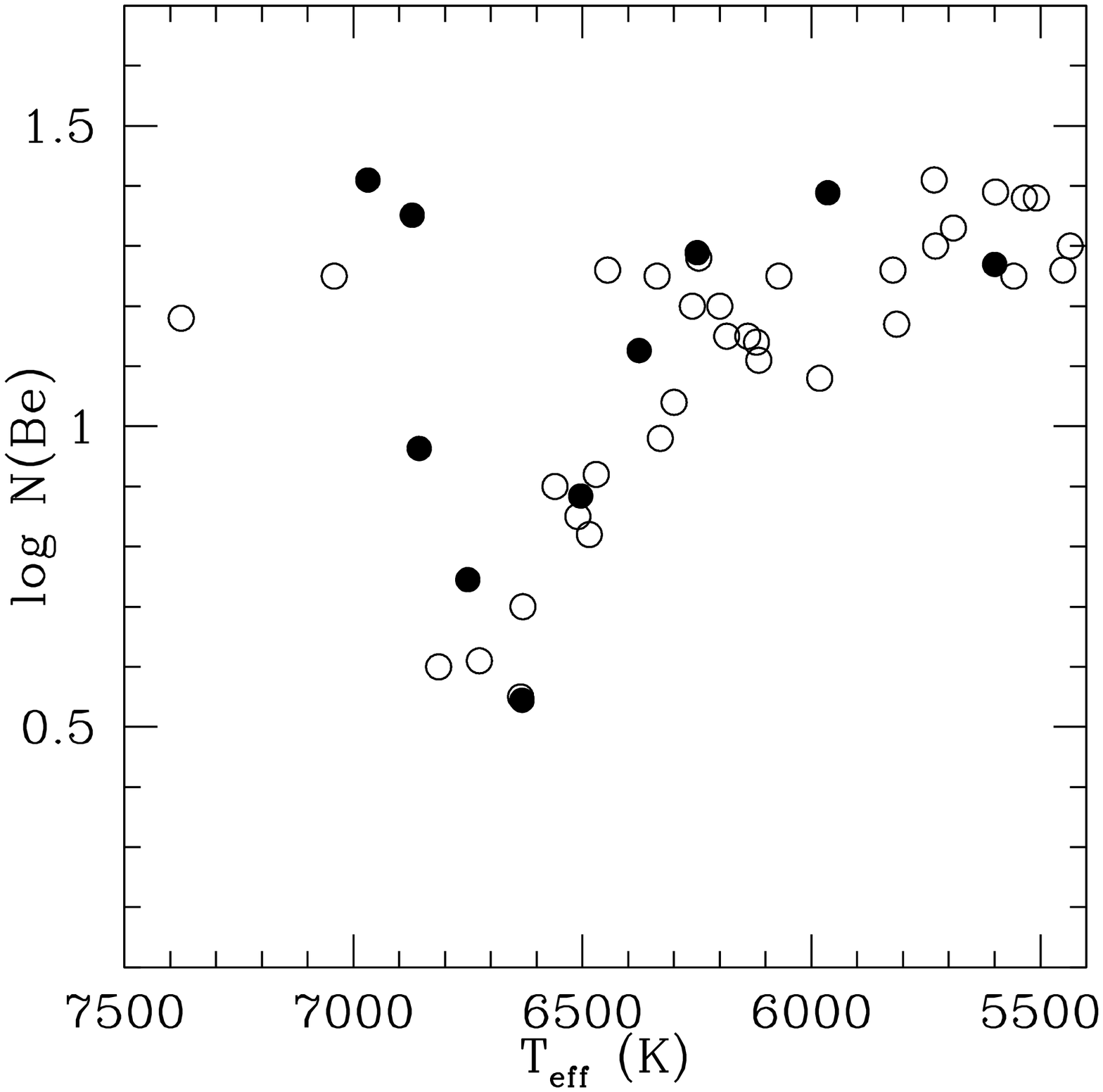}
\caption{Comparison between the computations of lithium (left) and beryllium (right) depletion and the observational values. The solid circles represents the models computed in these paper. Left figure : the open hexagons are observations from Boesgaard \& Tripicco (1986), the open stars are from Boesgaard \& Budge (1988), the open squares are from Thorburn et al (1993) and the open circles are observations from Boesgaard \& Kind (2002), the inverted triangles denote upper limits. Right figure : the open circles are from Boesgaard \& Kind (2002).}
\label{obs}
\end{figure}

The result obtained in this paper are compared to the observational points given by Boesgaard \& King (2002), Thorburn et al. (1993), Boesgaard \& Budge (1988), Boesgaard \& Tripicco (1986).
The original lithium and beryllium abundances are the meteoritic ones : A(Li)=3.31 dex and A(Be)=1.42 dex, both values are from Anders \& Grevesse (1989). Figure \ref{obs} shows the obtained results. The open symbols represent the observational points. The solid ones represent our models. 

All the explanations which have been proposed up to now to account for the ``lithium dip" in the Hyades have arbitrary parameters which are adjusted at some point. For this reason the strength of a theory or a modelisation does not really rely on the fact that the lithium observations are well reproduced. Other fits and signatures of the invoked processes have to be found to test the theory.

We do not escape the rule : as pointed out in section 5.1, the diffusion coefficient parameters have been adjusted to obtain the best fit with the observations for lithium in the Hyades. However, once this has been done, there is no arbitrary parameter for the fit of the beryllium abundances, which is excellent. The results are thus very encouraging. Furthermore, we stress that the effective temperature for which the meridional circulation changes from one cell to two different cells is precisely that of the lithium dip, as already pointed out by CVZ92. This fact is independent from diffusion coefficient parameters. It can only slightly vary according to the prescription of the computations of the convective zone depth (here the mixing length theory, computed as for the Sun). 

We claim that the dip feature is indeed related to this separation of the meridional circulation in two different cells. We have shown here that the two cells can be well disconnected due to the process of ``creeping paralysis" which occurs at the top of the deepest cell, thereby increasing the thickness of the frozen zone in-between. Consequences on the lithium abundances in sub-giants originating from stars hotter than the dip will be studied in a forthcoming paper (Th\'eado, Vauclair and Dias do Nascimento 2003).

\end{document}